\documentclass[12pt]{article}
\usepackage{amsthm} 
\usepackage{amsmath} 
\usepackage{amsfonts}
\usepackage{amsxtra}
\usepackage{yhmath}
\usepackage{amssymb} 
\usepackage{mathrsfs} 
\usepackage{esint} 
\usepackage{empheq}
\usepackage{stmaryrd}
\usepackage{dsfont}
\usepackage[latin1]{inputenc}
\usepackage{upgreek}
\usepackage[all]{xy}
\usepackage{graphicx}
\usepackage{latexsym}
\usepackage{color}
\usepackage{epsfig}
\usepackage{graphicx}
\usepackage{caption}
\usepackage{subcaption}
\usepackage{here}
\usepackage{float}
\usepackage{array}
\usepackage{tikz}
\usepackage{tikz-cd}
\usetikzlibrary{3d,arrows,calc,chains,matrix,positioning,scopes}

\DeclareSymbolFont{UPM}{U}{eur}{m}{n}
\DeclareMathSymbol{\uppartial}{0}{UPM}{"40}

\def\ba{\begin{array}}
\def\ea{\end{array}}

\def\bea{\begin{eqnarray}}
\def\eea{\end{eqnarray}}

\newcommand{\egzz}{\, \stackrel{\mathbb{Z}}{{=}}\,}

\newcommand{\oppartial}{\mathop{}\mathopen{}\uppartial}

\newcommand{\image}{\mathop{}\mathopen{}\mathrm{Im}\,}
\newcommand{\noyau}{\mathop{}\mathopen{}\mathrm{Ker}\,}

\newcommand{\ensemblenombre}[1]{\mathbb{#1}}

\newcommand{\Z}{\ensemblenombre{Z}}

\newcommand{\R}{\ensemblenombre{R}}
\newcommand{\C}{\ensemblenombre{C}}

\newcommand{\intervalle}[4]{\mathopen{#1}#2\mathclose{}\mathpunct{};#3\mathclose{#4}}

\newcommand{\intervalleentier}[2]{\intervalle{\llbracket}{#1}{#2}{\rrbracket}}

\newcommand{\slfrac}[2]{\left.#1\middle/#2\right.}

\newcommand{\pgcd}[2]{\mathrm{\textsc{gcd}}\mathopen{}\left(#1,#2\right)}

\newcommand{\link}[2]{\mathrm{{\ell}k}\mathopen{}\left(#1,#2\right)}

\newcommand{\abs}[1]{\left\lvert#1\right\rvert}

\newcommand{\Hom}[2]{\mathrm{Hom}\mathopen{}\left(#1,#2\right)}

\theoremstyle{definition}

\theoremstyle{plain}

\newtheorem*{lem}{Lemma}
\newtheorem*{dem}{Proof}

\begin{document}

\pagestyle{empty}
\setcounter{page}{0}
\hspace{-1cm}

\begin{center}
{\Large {\bf A reciprocity formula from abelian BF and Turaev-Viro theories}}%
\\[1.5cm]

{\large P. Mathieu and F. Thuillier}

\end{center}

\vskip 0.7 truecm

{\it LAPTH, Université Savoie Mont Blanc, CNRS, 9, Chemin de Bellevue, BP 110, F-74941
Annecy-le-Vieux cedex, France.}

\vspace{3cm}

\centerline{{\bf Abstract}}
In this article we show that the use of Deligne-Beilinson cohomology in the context of the $U(1)$ BF theory on a closed 3-manifold $M$ yields a discrete $\Z_N$ BF theory whose partition function is an abelian TV invariant of $M$. By comparing the expectation values of the $U(1)$ and $\mathbb{Z}_N$ holonomies in both BF theories we obtain a reciprocity formula.

\vspace{2cm}

\vfill
\newpage
\pagestyle{plain} \renewcommand{\thefootnote}{\arabic{footnote}}

\section{Introduction}

The impact of Deligne-Beilinson cohomology in the context of Quantum Field Theory was carefully investigated in \cite{BGST}. In a previous article \cite{MT} a study of the $U(1)$ BF theory within the Deligne-Beilinson cohomology \cite{De,Be} framework was initiated, following what was done in the $U(1)$ Chern-Simons (CS) theory case \cite{GT1,T1,GT2,GT3,T2}. In this first article the partition function of the BF theory was computed and compared with the absolute square of the Chern-Simons partition function thus highlighting significant differences from the non-abelian case. In this same article an abelian Turaev-Viro (TV) invariant, whose construction is based on a generalisation of V. Turaev and O. Viro \cite{TV} approach as proposed by B. Balsam and A. Kirillov \cite{BK}, was exhibited and it was shown that up to a normalisation this abelian TV invariant coincides with the $U(1)$ BF partition function.

In the second section of this article we complete the study of the $U(1)$ BF theory on a closed $3$-manifold $M$ by computing expectation values of $U(1)$ holonomies, still in the Deligne-Beilinson (DB) cohomology framework. In section 3 we show that the Turaev-Viro invariant can be seen as the partition function of a discrete $\Z_N$ BF theory whose observables are $\Z_N$ holonomies. Some gauge fixing procedures are also discussed in this section together with the usefulness of a Heegaard splitting of $M$. Finally by taking for $N$ the quantized coupling constant of the original $U(1)$ BF theory a relationship between expectation values of the BF and TV theories is made explicite in section 4. This yields a reciprocity formula which is comparable with Deloup-Turaev one \cite{DT}, this last formula being related to the $U(1)$ Chern-Simons theory and the Reshetekhin-Turaev surgery formula \cite{RT,GT3,T2,MT}.

The use of DB cohomology proves to be very effective in the $U(1)$ BF theory since unlike the non-abelian $SU(2)$ case we find that: 1) the discretisation of the original $U(1)$ BF theory is a consequence of the construction and not an input; 2) no regularisation of the expectation values is required in the discrete abelian case because all sums occurring are finite whereas a Quantum Group has to be introduced by hand in the non-abelian case to get well-defined expressions \cite{B,FL}.

The results obtained in this article can be gathered into:
\vspace{2mm}

{\it Proposition. For a smooth, closed, connected and oriented three-manifold $M$ endowed with dual cellular decompositions $\mathcal{C}$ and $\mathcal{C}^\ast$ we have:

(1) In the $U(1)$ BF theory the expectation values of the $U(1)$-holonomies along two cycles $\gamma_{1}$ and $\gamma_{2}$ are:
\bea
\begin{aligned}
\left\langle\left\langle \gamma_{1} , \gamma_{2} \right\rangle\right\rangle_{BF_{N}}
 = \; & \delta^{\left[N\right]}_{\textbf{f}_{1}}
\delta^{\left[N\right]}_{\textbf{f}_{2}} \, e^{-\frac{2i\pi}{N} \link{\gamma^{0}_{1} + \gamma^{\tau}_{1}}{\gamma^{0}_{2} + \gamma^{\tau}_{2}}}
\! \! \sum\limits_{{\boldsymbol{\kappa}}_{1},{\boldsymbol{\kappa}}_{2}\in T_{1}}
e^{-2i\pi \left( N Q\left({\boldsymbol{\kappa}}_{1},{\boldsymbol{\kappa}}_{2}\right)
+ Q\left({\boldsymbol{\kappa}}_{1} , {\boldsymbol{\tau}}_{1}\right)
+ Q\left({\boldsymbol{\kappa}}_{2} , {\boldsymbol{\tau}}_{2}\right) \right)}
\, ,
\end{aligned}
\eea
where $\gamma_{1} = \gamma^{0}_{1} + \gamma^{f}_{1} + \gamma^{\tau}_{1}$ and $\gamma_{1} = \gamma^{0}_{1} + \gamma^{f}_{1} + \gamma^{\tau}_{1}$ is a decomposition of these cycles into their trivial, free and torsion part, ${\textbf{f}_{1}}$ and ${\textbf{f}_{2}}$ denote the free homology classes and ${\boldsymbol{\tau}}_{1}$ and ${\boldsymbol{\tau}}_{2}$ the torsion homology classes of $\gamma_{1}$ and $\gamma_{2}$, and $Q$ is the linking form on torsion.

(2) There is an abelian TV theory whose observables are $\Z_N$-holonomies, the expectation values of which are defined by:
\bea
\left\langle\left\langle z_{1}, z_{2}\right\rangle\right\rangle_{TV_{N}} = \frac{1}{N^{F+V-1}}
\displaystyle\sum\limits_{{\textbf{m}}\in\Z^{F}_{N}}
\displaystyle\sum\limits_{{\textbf{l}}\in\Z^{E}_{N}}
e^{\frac{2i\pi}{N} \left( {\textbf{m}} \cdot d{\textbf{l}}
+ {\textbf{l}} \cdot {\textbf{z}}_{1}
+ {\textbf{m}}\cdot{\textbf{z}}_{2} \right) } \, ,
\eea
where $z_1$ and $z_2$ are two cycles of $\mathcal{C}$ and $\mathcal{C}^\ast$ respectively represented by ${\textbf{z}}_{1} \in \Z^E$ and ${\textbf{z}}_{2} \in \Z^F$, and with $F$, $E$ and $V$ the number of faces, edges and vertices of $\mathcal{C}$.

(3) The TV and BF observables expectation values satisfy:
\bea
\nonumber
\left\langle\left\langle z_{1}, z_{2}\right\rangle\right\rangle_{TV_{N}}
= \frac{N^{b_{1}}}{p_{1} \cdots p_{n}} \left\langle\left\langle z_{1} , z_{2} \right\rangle\right\rangle_{BF_{N}} \, ,
\eea
which provides a reciprocity formula.}
\vspace{2mm}

All along this article $M$ is a smooth, closed, connected and oriented three-manifold. We use $\egzz$ to denote equality in $\mathbb{R}/\mathbb{Z}$, that is to say modulo integers, as well as Einstein summation convention.

\section{Abelian BF theory}

\subsection{Reminders on Deligne-Beilinson cohomology}

We denote by $Z_p(M)$ the set of singular $p$-cycles in $M$ and by $H_p(M)$ (resp. $H^p(M)$) the corresponding homology (resp. cohomology) group. The space of smooth $p$-forms on $M$ is denoted by $\Omega^p(M)$, the subset of closed $p$-forms by $\Omega_0^p(M)$ and the one of closed $p$-forms with integral periods by $\Omega _{\Z}^1 \left( M \right)$. The Pontrjagin dual of $\Omega _{\Z}^p \left( M \right)$ is ${\Omega_{\Z}^p(M)}^{\ast} \equiv Hom\left( \Omega_{\Z}^p(M),{{\R}/{\Z}} \right)$ and the set of de Rham $p$-currents in $M$ is the (topological) dual of $\Omega _{\mathbb Z}^{3-p} \left( M \right)$. In particular, every $p$-chain $c$ in $M$ defines a de Rham $(3-p)$-current denoted by $j_c$.
Poincaré duality states that $H^2(M) \simeq H_1(M)$. Hence when referring to the class of a $1$-cycle in $M$ we indifferently refer to its homology class or to the cohomology class of its Poincaré dual. We use the canonical decomposition of the abelian group $H^2(M)$ into its free and torsion part according to: $\Z^{b_1} \oplus \Z_{p_1} \oplus \cdots \oplus \Z_{p_n}$, where $b_1$ is the first Betti number of $M$, and $p_i|p_{i+1} \in \Z$ for $i = 1, \cdots , n-1$.

As in the $U(1)$ Chern-Simons theory, the space of fields of the $U(1)$ BF theory is built from the first Deligne-Beilinson cohomology group of $M$, $H_D^1(M,\Z)$, or its Pontrjagin dual $H_D^1(M,\Z)^{\ast} \equiv Hom\left( {H_D^1 \left( {M,{\mathbb Z}} \right),{{\mathbb R}/{\mathbb Z}}} \right)$. This means in particular that we deal with classes of $U(1)$-connections rather than with connections.

The spaces $H_D^1(M,\Z)$ and $H_D^1(M,\Z)^{\ast}$ are $\Z$-modules. They can be embedded into the exact sequence:
\begin{equation}
\label{exactsequence1}
0 \longrightarrow {\Omega^1( M) \over {\Omega_{\Z}^1(M)}} \longrightarrow H_D^1(M,\Z)
\longrightarrow H^{2}(M,\Z) \longrightarrow 0 \, ,
\end{equation}
for the former and:
\begin{equation}
\label{exactsequence2}
0 \longrightarrow {\Omega_{\Z}^2(M)}^{\ast}  \longrightarrow H_D^1(M,\Z)^{\ast}
\longrightarrow H^{2}(M,\Z) \longrightarrow 0 \, ,
\end{equation}
for the latter. The space $\Omega _{\mathbb Z}^1 \left( M \right)$ is thus the global gauge group of smooth $U(1)$-connections on $M$.

The configuration space of the $U(1)$ BF theory is the product $H_D^1(M,\Z) \times H_D^1(M,\Z)$, or at the level of distributions $H_D^1(M,\Z)^{\ast} \times H_D^1(M,\Z)^{\ast}$.
\vspace{2mm}

Let us list some important properties of $H_D^1(M,\Z)$ and $H_D^1(M,\Z)^{\ast}$.
\vspace{2mm}

{\it (1) Regular DB classes.} Relating the two previous exact sequences by the mean of the canonical injection ${\Omega^1( M) / {\Omega_{\Z}^1(M)}} \rightarrow {\Omega_{\Z}^2(M)}^{\ast}$ we deduce that there is a canonical injection:
\bea
\label{inclusion}
H_D^1(M,\Z) \hookrightarrow H_D^1(M,\Z)^{\ast} \, .
\eea
Hence we can identify smooth DB classes as regular elements of $H_D^1(M,\Z) \rightarrow H_D^1(M,\Z)^{\ast}$ just like smooth functions are identified with regular distributions. Moreover there is a canonical injection of de Rham $1$-currents into ${\Omega_{\Z}^2(M)}^{\ast}$. For instance the de Rham currents of two surfaces with the same boundary define the same DB class. However, since there is no real possible confusion, we will use the same notation for a current and its image in ${\Omega_{\Z}^2(M)}^{\ast}$.
\vspace{2mm}

{\it (2) Bundle structure.} Exact sequence \eqref{exactsequence1} (resp. \eqref{exactsequence2}) tells us that $H_D^1(M,\Z)$ (resp. $H_D^1(M,\Z)^{\ast}$) is a bundle over the discrete set $H^{2}(M,\Z)$ whose fibres are affine spaces with associated vector space ${\Omega^1( M) / {\Omega_{\Z}^1(M)}}$ (resp. ${\Omega_{\Z}^2(M)}^{\ast}$). Hence for any $A \in H_D^1(M,\Z)$ and any $\omega \in {\Omega^1( M) / {\Omega_{\Z}^1(M)}}$ we write $A + \omega$ the DB class obtained from $A$ by the translation $\omega$.
The fiber over the zero class of $H^2(M)$ is called the \textbf{trivial fibre}.
A fibre over a purely free class of $H^2(M)$ is called a \textbf{free fibre}.
A fibre over a purely torsion class of $H^2(M)$ is called a \textbf{torsion fibre}.
\vspace{2mm}

{\it (3) DB product.}  There is a commutative product:
\begin{equation}
\label{DBproduct}
\star : H_D^1 \left( {M,{\mathbb Z}} \right) \times H_D^1 \left( {M,{\mathbb Z}}
\right) \longrightarrow {\Omega^3(M) \over {\Omega_{\Z}^3(M)}} \, .
\end{equation}
Composing this product with integration over $M$ provides a $\R/\Z$-valued symmetric linear pairing in $H_D^1(M,\Z)$:
\begin{equation}
\label{prodictduality}
\int_M \circ \; \, \star : H_D^1 \left( {M,{\mathbb Z}} \right) \times H_D^1 \left( {M,{\mathbb Z}}
\right) \longrightarrow \R/\Z \, .
\end{equation}
The DB product and the pairing can be straightforwardly extended to $H_D^1 \left( {M,{\mathbb Z}} \right) \times H_D^1 \left( {M,{\mathbb Z}}
\right)^{\ast}$.
\vspace{2mm}

{\it (4) Holonomy.} There is a pairing:
\begin{equation}
\label{integral}
\oint : H_D^1 \left( {M,{\mathbb Z}} \right) \times Z_1 \left( {M}
\right)\longrightarrow {{\mathbb R}/{\mathbb Z}} \, ,
\end{equation}
which defines integration of DB classes along cycles in $M$. From this pairing we deduce the inclusion:
\bea
\label{injcycles}
Z_1(M) \subset H_D^1(M,\Z)^{\ast} \, ,
\eea
which means that we can associate to a $1$-cycle $\gamma$ a unique DB class $\eta_\gamma \in H_D^1(M,\Z)^{\ast}$ defined by:
\begin{equation}
\label{dualcycle}
\forall A \in H_D^1(M,\Z) , \; \; \oint_\gamma A \egzz \int_M A \star \eta_\gamma \, .
\end{equation}
This pairing yields the holonomy of a DB class $A$ according to:
\begin{equation}
\label{holonomy}
e^{2 i \pi \oint_\gamma A } = e^{2 i \pi \int_M A \star \eta_\gamma} \, .
\end{equation}

{\it (5) Regularisation.} For the same reason as the product of distributions is ill-defined, some regularisation procedure has to be chosen to extend product \eqref{DBproduct} and pairing \eqref{prodictduality} to $H_D^1 \left( {M,{\mathbb Z}} \right)^{\ast} \times H_D^1 \left( {M,{\mathbb Z}}\right)^{\ast}$. For DB classes of $1$-cycles of $M$ we adopt the so-called \textbf{zero regularisation} convention which is defined by:
\begin{equation}
\label{zeroregul}
\int_M \eta_\gamma \star \eta_\gamma \egzz 0 \, .
\end{equation}
\vspace{2mm}

{\it (6) Origins.} The zero fibre admits as canonical origin the zero DB class which is the class of the zero $U(1)$ connection. This choice of origin allows to identify this fibre with the translation group ${\Omega^1( M) / {\Omega_{\Z}^1(M)}}$ (or ${\Omega_{\Z}^2(M)}^{\ast}$). On any other fibre of $H_D^1 \left( {M,{\mathbb Z}}\right)$ and $H_D^1 \left( {M,{\mathbb Z}}\right)^{\ast}$ there is no such canonical choice.

Even if there is no specific origin on free fibres, injection \eqref{injcycles} suggests the following in $H_D^1 \left( {M,{\mathbb Z}}\right)^{\ast}$: let $\varsigma_a$ ($a = 1, \cdots ,b_1$) be a set of once for all chosen $1$-cycles of $M$ which generate $F_1(M)$; then the DB class of each $1$-cycle $\sum m^a \varsigma_a$ is taken as origin of the fiber over $\boldsymbol{m} = (m^1, \cdots , m^{b_1}) \in F^2(M) \simeq F_1(M)$. Such origins will be referred as {\bf free origins} and denoted by $A^{\varsigma}_{\boldsymbol{m}}$. Note that $A^{\varsigma}_{\boldsymbol{m}} = \sum m^a \eta_{\varsigma_a}$ so that zero regularisation also applies to $A^{\varsigma}_{\boldsymbol{m}}$.

Although torsion $1$-cycles could also be chosen as origin on torsion fibres in $H_D^1 \left( {M,{\mathbb Z}}\right)^{\ast}$, there exists specific origins on these fibres for both $H_D^1 \left( {M,{\mathbb Z}}\right)$ and $H_D^1 \left( {M,{\mathbb Z}}\right)^{\ast}$. Indeed it can be shown \cite{GT3} that on each torsion fibre there exists a DB class, $A^c_{\boldsymbol{\tau}}$, such that:
\begin{equation}
\label{torsionorigin}
\int_M A^c_{\boldsymbol{\tau}_1} \star A^c_{\boldsymbol{\tau}_2} \egzz - Q(\boldsymbol{\tau}_1 , \boldsymbol{\tau}_2)  \; \; \; \mbox{ and } \; \; \;  \int_M A^c_{\boldsymbol{\tau}} \star \omega \egzz 0 \, ,
\end{equation}
for any $\omega \in {\Omega^1( M) / {\Omega_{\Z}^1(M)}}$ (or ${\Omega_{\Z}^2(M)}^{\ast}$), with $Q : T_1(M) \times T_1(M) \longrightarrow \R/\Z$ the linking form of $M$.
With some abuse such particular origins are called \textbf{canonical torsion origins}. Since for any representative $\tau$ of a torsion class $\boldsymbol{\tau}$ there is $p_{\boldsymbol{\tau}} \in \Z$ and a $2$-chain $\Sigma_{\tau}$ of $M$ such that $p_{\boldsymbol{\tau}} \tau = \partial \Sigma_{\tau}$ then we have:
\bea
\label{decomptorsion}
\eta_{\tau} = A^c_{\boldsymbol{\tau}} + {j_{\Sigma_{\tau}}/p}
\eea
and thus:
\begin{equation}
\label{linking}
\int_M \eta_{\tau} \star \eta_{\tau} \egzz \int_M A^c_{\boldsymbol{\tau}} \star A^c_{\boldsymbol{\tau}} + 2 \int_M A^c_{\boldsymbol{\tau}}  \star  {j_{\Sigma_{\tau}} \over p} +  \int_M {j_{\Sigma_{\tau}} \over p}  \star  {j_{\Sigma_{\tau}} \over p}   \, .
\end{equation}
Let us recall that even if $\Sigma'_{\tau}$ is another $2$-chain such that $p_{\boldsymbol{\tau}} \tau = \partial \Sigma'_{\tau}$ we have ${j_{\Sigma_{\tau}} / p} = {j_{\Sigma'_{\tau}} / p}$ in ${\Omega_{\Z}^2(M)}^{\ast}$. Using relations \eqref{zeroregul} and  \eqref{torsionorigin} we find that:
\begin{equation}
\label{linking2}
\int_M {j_{\Sigma_{\boldsymbol{\tau}}} \over p}  \star  {j_{\Sigma_{\boldsymbol{\tau}}} \over p} \egzz Q(\boldsymbol{\tau} , \boldsymbol{\tau}) \egzz \int_M {j_{\Sigma_{\boldsymbol{\tau}}} \over p} \wedge  d  {j_{\Sigma'_{\boldsymbol{\tau}}} \over p} \, ,
\end{equation}
which shows the consistency of the construction.

A generic DB class $A \in H_D^1 \left( {M,{\mathbb Z}} \right)^{\ast}$ is decomposed according to:
\bea
\label{gendecompos}
A = A^{\varsigma}_{{\textbf{m}}} + A^c_{{\boldsymbol{\kappa}}} + \alpha \, ,
\eea
with $\alpha \in {\Omega_{\Z}^2(M)}^{\ast}$.
\vspace{2mm}

{\it (7) Zero modes.} The set ${\Omega^1( M) / {\Omega_{\Z}^1(M)}}$ of translations in $H_D^1(M,\Z)$ can be embedded on its turn into an exact sequence:
\begin{equation}
\label{exactsequence3}
0 \longrightarrow {\Omega_0^1(M) \over {\Omega_{\Z}^1(M)}} \longrightarrow {\Omega^1(M) \over {\Omega_{\Z}^1(M)}} \longrightarrow {\Omega^1(M) \over \Omega_0^1(M)} \longrightarrow 0 \, ,
\end{equation}
where $\Omega_0^1(M)$ denotes the space of closed 1-forms on $M$ (see details in \cite{GT2,GT3}). This implies that we can (non-canonically) write:
\begin{equation}
\label{prodzeromods}
\left({\Omega^1(M) \over {\Omega_{\Z}^1(M)}}\right) \simeq \left({\Omega^1(M) \over \Omega_0^1(M)}\right) \times \left({\Omega_0^1(M) \over {\Omega_{\Z}^1(M)}}\right)  \, .
\end{equation}
Elements of ${\Omega_0^1(M) / {\Omega_{\Z}^1(M)}}$ are called \textbf{zero modes}. It is obvious that:
\begin{equation}
\label{zeromods}
\left({\Omega_0^1(M) \over {\Omega_{\Z}^1(M)}}\right) \simeq \left(\R \over \Z\right)^{b_1} \, .
\end{equation}
We refer to this quotient as the space of zero modes. It can be shown \cite{GT2,GT3} that for any zero-mode $\alpha_0$:
\begin{equation}
\label{isotropy2}
\forall \omega \in {\Omega_{\Z}^2(M)}^{\ast}, \quad \int_M \alpha_0 \star \omega \egzz 0 \, .
\end{equation}
By combining the second equation of \eqref{torsionorigin} together with decomposition \eqref{gendecompos} and property \eqref{isotropy2} we find that:
\bea
\label{onlyfree}
\forall A \in H_D^1(M,\Z)^{\ast}, \quad \int_M \alpha_0 \star A \egzz \int_M \alpha_0 \star A^{\varsigma}_{{\textbf{m}}} \egzz \sum_{a = 1}^{b_1} m^a \oint_{\varsigma_a} \alpha_0 \, .
\eea

Let us consider a set of smooth closed $1$-forms $\rho_a$ ($a = 1, \cdots , b_1$) such that:
\bea
\label{dualform}
\oint_{\varsigma_a} \rho_b = \delta_{ab} \, ,
\eea
the $\varsigma_a$'s being the $1$-cycles defining the free origins $A^\varsigma_{\boldsymbol{m}}$. The images in ${\Omega_0^1(M) / {\Omega_{\Z}^1(M)}}$ of these 1-forms form a basis $\left\{\rho_a\right\}_{a=1,\dots,b_1}$ for zero-modes according to:
\bea
\label{basiszeromod}
\alpha_0 = \theta^a \rho_a  \, ,
\eea
with $\theta^a \in \mathbb{R/\mathbb{Z}}$. The components $\theta^a$ depends on the zero mode $\alpha_0$ and not on the basis $\left\{\rho_a\right\}_{a=1,\dots,b_1}$. We have:
\bea
\label{scalzeromod}
\int_M A^{\varsigma}_{\boldsymbol{m}} \star (\theta^b \rho_b) \egzz \sum_{a,b = 1}^{b_1} m^a \theta^b \int_{\varsigma_a} \rho_b = \sum_{a = 1}^{b_1} m^a \theta^a \equiv \boldsymbol{m} \cdot \boldsymbol{\theta} \, .
\eea
By definition the closed $1$-forms $\rho_a$ are the Poincaré dual of some closed surfaces $S^0_a$ in $M$ which generate $F_2(M) = H_2(M)$. This means that instead of $\rho_a$ we could use the de Rham currents $j_{S^0_a}$ of these surfaces, relation \eqref{dualform} then becoming the intersection number of $S^0_b$ and $\varsigma_a$. This means that the splitting of $\left({\Omega^1(M) / {\Omega_{\Z}^1(M)}}\right)$ straightforwardly extends to ${\Omega_{\Z}^2(M)}^{\ast}$. The decomposition of $\alpha \in {\Omega_{\Z}^2(M)}^{\ast}$ according to this splitting is written:
\bea
\label{splittingmode}
\alpha = \alpha_0 + \alpha_{\perp} \, .
\eea
In fact it is more rigorous to say that $\alpha_0 + \alpha_{\perp}$ biunivocally span ${\Omega_{\Z}^2(M)}^{\ast}$ when $\alpha_0$ runs trough $\Omega_0^1(M) / {\Omega_{\Z}^1(M)}$ and $\alpha_{\perp}$ through $\Omega^1(M) / \Omega_0^1(M)$ (or its distributional version).
Finally any $A \in H_D^1 \left( {M,{\mathbb Z}} \right)^{\ast}$ is decomposed as:
\bea
\label{gendecomposfinal}
A = A^{\varsigma}_{{\textbf{m}}} + A^c_{{\boldsymbol{\kappa}}} + \alpha_0 + \alpha_{\perp} \, .
\eea

\subsection{Abelian BF action, measure, partition function and observables}

Locally, i.e. in any open set diffeomorphic to $\R^{3}$, $A \star B = A \wedge dB$, which is the lagrangian usually considered in the $U\left(1\right)$ BF theory. This suggests to set:
\bea
\label{BFaction}
BF_{N}\left(A,B\right) = \int_{M}\mathfrak{bf}_{N}\left(A,B\right) = N \int_{M} A \star B \, ,
\eea
as generalized $U(1)$ BF action with coupling constant $N$, where $(A,B) \in H^{1}_{D} \times H^{1}_{D}$.
From pairing \eqref{prodictduality} we deduce that $BF_{N}\left(A,B\right)$ is well defined if and only if $N \in \Z$.

At the quantum level we assume that our gauge fields live in a configuration space $\mathcal{H} \subset \left(H^{1}_{D}\right)^{*}$ which contains $H^{1}_{D}$ and $Z_{1} \times Z_{1}$ so that $(A,B) \in \mathcal{H}^2 = \mathcal{H} \times \mathcal{H} $. In particular $\mathcal{H}$  has an affine bundle structure over $H^2(M)$ whose translation group $\mathcal{T} \subset {\Omega_{\Z}^2(M)}^{\ast}$. We also assume that the set of free and torsion origins previously discussed, $A^0_{\boldsymbol{m}}$ and $A^c_{\boldsymbol{\tau}}$, has been set on $\mathcal{H}$.

We provide $\mathcal{H}^2$ with the (formal) measure $d \mu_{N}$ defined by:
\bea
\forall \left(A,B\right) \in \mathcal{H}^2 , \quad d \mu_{N}\left(A,B\right) = \mathcal{D}\!A \, \mathcal{D}\!B \, e^{2i\pi BF_{N}\left(A,B\right)} \, ,
\eea
where $\mathcal{D}$ stands for the (formal) Lebesgue measure on $\mathcal{H}$. The measure $d \mu_{N}$ satisfies the fundamental property:
\bea
\label{measurelinear}
d\mu_{N} \left( A + A_0 ,B + B_0 \right)
= d\mu_{N} \left( A,B \right)
e^{2i\pi \left\{ BF_{N}\left(A_0,B\right)
+ BF_{N}\left(A,B_0\right)
+ BF_{N}\left(A_0,B_0\right) \right\}} \, ,
\eea
for any fixed $(A_0,B_0) \in \mathcal{H}^2$. This means that, unlike $\mathcal{D}$, the measure $d \mu_{N}$ is not invariant by translation. However it has an invariance associated with zero modes. Consider $j_{1}$ and $j_{2}$ the de Rham currents of two closed surfaces $\Sigma_{1}$ and $\Sigma_{2}$ in $M$. Then $j_{1}$ and $j_{2}$ are zero modes and for all $\left(m,n\right) \in \Z^{2}$ and all $\left(A,B\right) \in \mathcal{H}^2$, properties \eqref{scalzeromod} and \eqref{measurelinear} imply that:
\bea
d\mu_{N}\left(A + m \frac{j_{1}}{N},B + n \frac{j_{2}}{N}\right)
= d \mu_{N}\left(A,B\right) \, .
\eea

The BF partition function for a given coupling constant $N$ is defined as:
\bea
\label{BFparitiondef}
Z_{BF_{N}} = \frac{1}{\mathcal{N}_{N}} \int_{\mathcal{H}^2}
\mathcal{D}\! A \, \mathcal{D}\! B \, e^{2i\pi BF_{N}\left(A,B\right)} \, ,
\eea
with:
\bea
\label{BFnormalisation}
\mathcal{N}_{N} = \int_{\mathcal{T}^2}
\mathcal{D}\alpha \, \mathcal{D}\beta\, e^{2i\pi BF_{N}\left(\alpha,\beta\right)}.
\eea

The observables for this theory are the $U(1)$ holonomies, also called Wilson loops, that is to say:
\bea
W\left(A,\gamma_{1},B,\gamma_{2}\right) = e^{2i\pi\oint_{\gamma_{1}}A}e^{2i\pi\oint_{\gamma_{2}}B}.
\eea

The expectation values are computed through the formula:
\bea
\label{vevBF}
\begin{aligned}
\left\langle\left\langle\gamma_{1},\gamma_{2}\right\rangle\right\rangle_{BF_{N}}
& = \left\langle\left\langle W\left(A,\gamma_{1},B,\gamma_{2}\right) \right\rangle\right\rangle\\
& = \frac{1}{\mathcal{N}_{N}} \int_{\mathcal{H}^2} d \mu_{N}\left(A,B\right)
e^{2i\pi\oint_{\gamma_{1}}A}e^{2i\pi\oint_{\gamma_{2}}B}
\end{aligned} \, .
\eea
We use the notation $\left\langle\left\langle\cdot,\cdot\right\rangle\right\rangle_{BF_{N}}$ to emphasize the fact that we are working with a particular normalization: usually $\mathcal{N}_{N}$ is chosen so that $\left\langle 0,0 \right\rangle_{BF_{N}} = 1$ while here $\left\langle\left\langle 0,0 \right\rangle\right\rangle_{BF_{N}} = Z_{BF_{N}}$.

It can be checked that the expectation value $\left\langle\left\langle \gamma_{1},\gamma_{2}\right\rangle\right\rangle_{BF_{N}}$ is $N$ nilpotent, that is to say:
\bea
\label{NNilpotency}
\left\{ \begin{aligned}
& \left\langle\left\langle N \gamma_{1} , \gamma_{2} \right\rangle\right\rangle_{BF_{N}} = \left\langle\left\langle 0,\gamma_{2} \right\rangle\right\rangle_{BF_{N}} \\
& \left\langle\left\langle \gamma_{1},N\gamma_{2} \right\rangle\right\rangle_{BF_{N}} = \left\langle\left\langle \gamma_{1},0 \right\rangle\right\rangle_{BF_{N}}
\end{aligned} \right.
\eea

\subsection{Computation of expectation values}

Consider $\gamma_{1} = \gamma^{0}_{1} + \gamma^{f}_{1} + \gamma^{\tau}_{1}$ and $\gamma_{2} = \gamma^{0}_{2} + \gamma^{f}_{2} + \gamma^{\tau}_{2}$ where the superscript $0$ refers to the homologically trivial part of the loop, $f$ to its non-trivial free part and $\tau$ to its non-trivial torsion part.

If $\gamma^{f}_{1}$ is $N$ times a generator of the free part of the homology, then thanks to property \eqref{NNilpotency}:
\bea
\left\langle\left\langle \gamma_{1} = \gamma^{0}_{1} + \gamma^{f}_{1} + \gamma^{\tau}_{1} , \gamma_{2} \right\rangle\right\rangle_{BF_{N}}
= \left\langle\left\langle \gamma^{0}_{1} + \gamma^{\tau}_{1} , \gamma_{2}\right\rangle\right\rangle_{BF_{N}}.
\eea
and the same for $\gamma^{f}_{2}$. If not, given any closed surface $\Sigma$ and any integer $m$, we can write, together with the measure invariance:
\bea
\begin{aligned}
\left\langle\left\langle W\left(A+m\frac{j_{\Sigma}}{N},\gamma_{1},B,\gamma_{2}\right) \right\rangle\right\rangle
& = \left\langle\left\langle W\left(A,\gamma_{1},B,\gamma_{2}\right) \right\rangle\right\rangle\,
e^{2i\pi m \oint_{\gamma_{1}}\frac{j_{\Sigma}}{N}}\\
& = \left\langle\left\langle W\left(A,\gamma_{1},B,\gamma_{2}\right) \right\rangle\right\rangle \, .
\end{aligned}
\eea
Thus, if $\left\langle\left\langle W\left(A,\gamma_{1},B,\gamma_{2}\right) \right\rangle\right\rangle \neq 0$, we must have for any $m$:
\bea
e^{2i\pi m \oint_{\gamma_{1}}\frac{j_{\Sigma}}{N}} = 1,
\eea
which means that the intersection number of $\gamma^{f}_{1}$ and $\Sigma$ is $0$ modulo $N$ for all closed surface $\Sigma$. This contradicts the hypothesis that $\gamma^{f}_{1}$ is non-trivially free. Hence, $\left\langle\left\langle W\left(A,\gamma_{1},B,\gamma_{2}\right) \right\rangle\right\rangle$ must be $0$. The same reasoning apply to $B$ thus yielding:
\bea
\left\langle\left\langle\gamma_{1},\gamma_{2}\right\rangle\right\rangle_{BF_{N}}
= \delta_{\textbf{f}_{1}}^{\left[N\right]} \delta_{\textbf{f}_{2}}^{\left[N\right]} \left\langle\left\langle\gamma^{0}_{1} + \gamma^{\tau}_{1},\gamma^{0}_{2} + \gamma^{\tau}_{2}\right\rangle\right\rangle_{BF_{N}}
\, ,
\eea
where $\textbf{f}_{1}$ and $\textbf{f}_{2}$ denoting the homology class of $\gamma^{f}_{1}$ and $\gamma^{f}_{2}$, and with:
\bea
\delta_{\textbf{f}}^{\left[N\right]} = \left\{ \begin{aligned}
& 1 \quad \mbox{if}  \, \, \, \textbf{f} = 0 \, \, \, \mbox{modulo} \, \, N \\
& 0 \quad \mbox{otherwise} \, .
\end{aligned} \right.
\eea

We thus consider now:
\bea
\left\langle\left\langle\gamma^{0}_{1} + \gamma^{\tau}_{1},\gamma^{0}_{2} + \gamma^{\tau}_{2}\right\rangle\right\rangle_{BF_{N}}
= \frac{1}{\mathcal{N}_{N}} \int_{\mathcal{H}^2} d \mu_{N}\left(A,B\right) e^{2i\pi \oint_{\gamma^{0}_{1}+\gamma^{\tau}_{1}}A}
e^{2i\pi \oint_{\gamma^{0}_{2}+\gamma^{\tau}_{2}}B} \, .
\eea
Dualizing the loops of integration with Deligne classes $\eta_{0}$ associated to the trivial part $\gamma_{0}$ and $\eta_{\tau}$ associated to the torsion part $\gamma_{\tau}$ we can write:
\bea
\left\langle\left\langle\gamma^{0}_{1} + \gamma^{\tau}_{1},\gamma^{0}_{2} + \gamma^{\tau}_{2}\right\rangle\right\rangle_{BF_{N}}
= \frac{1}{\mathcal{N}_{N}} \int_{\mathcal{H}^{2}} \, d \mu_{N}\left(A,B\right)
e^{2i\pi \int_{M} \left\{ A \star \left(\eta^{0}_{1} + \eta^{\tau}_{1}\right) + B \star \left(\eta^{0}_{2}+\eta^{\tau}_{2}\right) \right\} } \, .
\eea
and using decomposion \eqref{decomptorsion} the right-hand side of the previous expression reads:
\bea
\label{3.40}
\frac{1}{\mathcal{N}_{N}} \int_{\mathcal{H}^{2}} \, d \mu_{N}\left(A,B\right) e^{2i\pi\int_{M} \left\{ A \star \left(\eta^{0}_{1} + A^c_{{\boldsymbol{\tau}}_{1}} + \frac{j_{\Sigma_{1}}}{p_{1}}\right) + B \star \left(\eta^{0}_{2} + A^c_{{\boldsymbol{\tau}}_{2}} + \frac{j_{\Sigma_{2}}}{p_{2}}\right) \right\} } \, .
\eea
By performing in this last expression the change of variables:
\bea
A \longrightarrow A-\left(\frac{\eta^{0}_{2}}{N}+\frac{j_{\Sigma_{2}}}{p_{2}N}\right)
\qquad
B \longrightarrow B-\left(\frac{\eta^{0}_{1}}{N}+\frac{j_{\Sigma_{1}}}{p_{1}N}\right) \, ,
\eea
expression \eqref{3.40} takes the form:
\bea
\frac{e^{-\frac{2i\pi}{N}\link{\gamma^{0}_{1} + \gamma^{\tau}_{1}}{\gamma^{0}_{2} + \gamma^{\tau}_{2}}}}{\mathcal{N}_{N}} \int_{\mathcal{H}^{2}} \, d \mu_{N}\left(A,B\right)
e^{2i\pi \int_{M} \left\{A \star A^c_{{\boldsymbol{\tau}}_{1}} + B \star A^c_{{\boldsymbol{\tau}}_{2}}\right\} } \, .
\eea

Using decomposition \eqref{gendecomposfinal} we get:
\bea
\left\{ \begin{aligned}
& A = A^{\varsigma}_{{\textbf{m}}_{1}} + A^c_{{\boldsymbol{\kappa}}_{1}} + \alpha_{0} + \alpha_{\perp} \\
& B = A^{\varsigma}_{{\textbf{m}}_{2}} + A^c_{{\boldsymbol{\kappa}}_{2}} + \beta_{0} + \beta_{\perp}
\end{aligned} \right. \, .
\eea
Contributions $A^{\varsigma}_{{\textbf{m}}_{1}} \star A^{\varsigma}_{{\textbf{m}}_{2}}$ cancel by zero regularisation whereas contributions $A^c_{{\boldsymbol{\kappa}}_{1}} \star \beta_{0}$, $\beta_{0} \star A^c_{{\boldsymbol{\tau}}_{2}}$, $A^c_{{\boldsymbol{\kappa}}_{2}} \star \alpha_{0}$, $\alpha_{0} \star A^c_{{\boldsymbol{\tau}}_{1}}$, $A^c_{{\boldsymbol{\kappa}}_{1}} \star \beta_{\perp}$, $\beta_{\perp} \star A^c_{{\boldsymbol{\tau}}_{2}}$, $A^c_{{\boldsymbol{\kappa}}_{2}} \star \alpha_{\perp}$, $\alpha_{\perp} \star A^c_{{\boldsymbol{\tau}}_{1}}$, $\alpha_{0} \star \beta_{\perp}$, $\alpha_{\perp} \star \beta_{0}$ et $\alpha_{0} \star \beta_{0}$ cancel thanks to properties \eqref{torsionorigin} and \eqref{isotropy2}. Thus, the only non trivial contributions are:
\bea
\left\{ \begin{aligned}
& N \int_{M} A^{\varsigma}_{{\textbf{m}}_{1}} \star A^c_{{\boldsymbol{\kappa}}_{2}} + A^{\varsigma}_{{\textbf{m}}_{2}} \star A^c_{{\boldsymbol{\kappa}}_{1}} +
A^{\varsigma}_{{\textbf{m}}_{1}} \star \beta_{0} + A^{\varsigma}_{{\textbf{m}}_{2}} \star \alpha_{0} \\
& N \int_{M} A^{\varsigma}_{{\textbf{m}}_{1}} \star \beta_{\perp} + A^{\varsigma}_{{\textbf{m}}_{2}} \star \alpha_{\perp}+
A^c_{{\boldsymbol{\kappa}}_{1}} \star A^c_{{\boldsymbol{\kappa}}_{2}} + \alpha_{\perp} \star \beta_{\perp} \\
& \int_{M} A^{\varsigma}_{{\textbf{m}}_{1}} \star A^c_{{\boldsymbol{\tau}}_{1}} + A^c_{{\boldsymbol{\kappa}}_{1}} \star A^c_{{\boldsymbol{\tau}}_{1}} \\
& \int_{M} A^{\varsigma}_{{\textbf{m}}_{2}} \star A^c_{{\boldsymbol{\tau}}_{2}} + A^c_{{\boldsymbol{\kappa}}_{2}} \star A^c_{{\boldsymbol{\tau}}_{2}}
\end{aligned} \right. \, .
\eea
We factorize out:
\bea
\left(\int\mathcal{D}\alpha_{0}e^{2i\pi N \int_{M} A^{\varsigma}_{{\textbf{m}}_{2}} \star \alpha_{0}}\right) \left(\int\mathcal{D}\beta_{0} e^{2i\pi N \int_{M} A^{\varsigma}_{{\textbf{m}}_{1}} \star \beta_{0}}\right) \, ,
\eea
and use relation \eqref{scalzeromod} to obtain:
\bea
\begin{aligned}
\sum \limits_{{\textbf{m}}_{1}\in F_{1}} \int \mathcal{D} \alpha_{0} e^{2i\pi N \int_{M} A^{\varsigma}_{{\textbf{m}}_{1}} \star \alpha_{0}}
& = \sum \limits_{{\textbf{m}}_{1}\in F_{1}} \left(
\prod\limits_{a=1}^{b_1} \int_{\R/\Z} d\theta^{i} e^{2i\pi N {m}^{a}_{1} \theta^{b} \int_{{\varsigma}_{a}}\rho_{b}} \right) \\
& = \sum \limits_{{\textbf{m}}_{1}\in F_{1}} \delta_{{\textbf{m}}_{1},0} \, .
\end{aligned}
\eea
Similarly we have:
\bea
\displaystyle\sum\limits_{{\textbf{m}}_{2}\in F_{1}}\int\mathcal{D}\beta_{0} e^{2i\pi N \int_{M}B_{{\textbf{m}}_{2}} \star \beta_{0}}
=\displaystyle\sum\limits_{{\textbf{m}}_{2}\in F_{1}} \delta_{{\textbf{m}}_{2},0} \, .
\eea
Our expectation value thus takes the form:
\bea
\left\langle\left\langle\gamma^{0}_{1} + \gamma^{\tau}_{1},\gamma^{0}_{2} + \gamma^{\tau}_{2}\right\rangle\right\rangle_{BF_{N}} =
\frac{e^{-\frac{2i\pi}{N}\link{\gamma^{0}_{1} + \gamma^{\tau}_{1}}{\gamma^{0}_{2} + \gamma^{\tau}_{2}}}}{\mathcal{N}_{N}} \times \; \; \; \; \; \; \; \; \; \; \; \; \; \; \; \; \; \; \; \; \; \; \; \; \; \; \; \; \; \; \; \; \; \; \; \; \; \; \; \; \; \; \; \; \; \;\; \; \; \;   \\ \nonumber  \times \sum\limits_{{\boldsymbol{\kappa}}_{1},{\boldsymbol{\kappa}}_{2}\in T_{1}}
\int \mathcal{D}\alpha_{\perp} \, \mathcal{D}\beta_{\perp}
 e^{2i\pi \int_{M} \left\{ N A^c_{{\boldsymbol{\kappa}}_{1}} \star A^c_{{\boldsymbol{\kappa}}_{2}} + \alpha_{\perp} \star \beta_{\perp} + A^c_{{\boldsymbol{\kappa}}_{1}} \star A^c_{{\boldsymbol{\tau}}_{1}} +
 A^c_{{\boldsymbol{\kappa}}_{2}} \star A^c_{{\boldsymbol{\tau}}_{2}} \right\} }.
\eea
In the same spirit, by factorising out the zero modes contribution in the expression of $\mathcal{N}_{N}$, we obtain:
\bea
\nonumber
\mathcal{N}_{N}
=\int\mathcal{D}\alpha_{\perp}\mathcal{D}\beta_{\perp} e^{2i\pi N \int_{M}\alpha_{\perp} \star \beta_{\perp}} \, ,
\eea
and since $\int_{M} A^c_{{\boldsymbol{\kappa}}_{1}} \star A^c_{{\boldsymbol{\kappa}}_{2}} = -Q\left({\boldsymbol{\kappa}}_{1},{\boldsymbol{\kappa}}_{2}\right)$ we finally have:
\bea
\label{finaldiscBF}
\left\langle\left\langle\gamma_{1},\gamma_{2}\right\rangle\right\rangle_{BF_{N}}
= \delta_{\textbf{f}_{1}}^{\left[N\right]} \delta_{\textbf{f}_{2}}^{\left[N\right]} e^{-\frac{2i\pi}{N}\link{\gamma^{0}_{1}+\gamma^{\tau}_{1}}{\gamma^{0}_{2}+\gamma^{\tau}_{2}}} \times  \; \; \; \; \; \; \; \; \; \; \;  \; \; \; \; \; \; \; \; \; \; \;  \; \; \; \; \; \; \; \; \; \; \;  \; \; \; \; \; \; \; \\ \nonumber
\times \sum\limits_{{\boldsymbol{\kappa}}_{1},{\boldsymbol{\kappa}}_{2}\in T_{1}}
e^{-2i\pi \left\{ N Q\left({\boldsymbol{\kappa}}_{1},{\boldsymbol{\kappa}}_{2}\right)
+ Q\left({\boldsymbol{\tau}}_{1} , {\boldsymbol{\kappa}}_{1}\right)
+ Q\left({\boldsymbol{\tau}}_{2} , {\boldsymbol{\kappa}}_{2}\right)\right\}} \, ,
\eea
which is the announced result. Note that we recover that \cite{MT}:
\bea
\left\langle\left\langle 0,0\right\rangle\right\rangle_{BF_{N}} = Z_{BF_{N}} = \sum\limits_{{\boldsymbol{\kappa}}_{1},{\boldsymbol{\kappa}}_{2}\in T_{1}}
e^{-2i\pi N Q\left({\boldsymbol{\kappa}}_{1},{\boldsymbol{\kappa}}_{2}\right)} \, .
\eea
Furthermore, if $M$ has no torsion the linking form $Q$ is trivial and hence \cite{GT1}:
\bea
\left\langle\left\langle\gamma,\gamma\right\rangle\right\rangle_{BF_{4k}}
= \left\langle\left\langle\gamma\right\rangle\right\rangle_{CS_{k}} \, .
\eea

\section{Towards an Abelian TV theory}

\subsection{Reminders on cellular decompositions}

We provide $M$ with an oriented cellular decomposition $\mathcal{C} = \left(\mathcal{P},\mathcal{F},\mathcal{E},\mathcal{V}\right)$ where $\mathcal{P}$ is the set of 3-cells (polyhedra), $\mathcal{F}$ the set of 2-cells (faces), $\mathcal{E}$ the set of 1-cells (edges) and $\mathcal{V}$ the set of 0-cells (vertices). These sets are given by:
\bea
\label{decompo}
\mathcal{P}=\left(P_{\mu}\right)_{\mu=1,\cdots,P} \;  ,  \;  \mathcal{F}=\left(S_{a}\right)_{a=1,\cdots,F} \; ,  \; \mathcal{E}=\left(e_{i}\right)_{i=1,\cdots,E} \;  ,  \; \mathcal{V}=\left(x_{\alpha}\right)_{\alpha=1,\cdots,V} \, .
\eea
As $M$ is closed, we can consider a dual oriented decomposition $\mathcal{C}^{*} = \left(\mathcal{P}^{*},\mathcal{F}^{*},\mathcal{E}^{*},\mathcal{V}^{*}\right)$ of $\mathcal{C}$ given by:
\bea
\label{dualdecompo}
\mathcal{V}^{*}=\left(x^{\mu}\right)_{\mu = 1,\cdots,P} \; , \; \mathcal{E}^{*}=\left(e^{a}\right)_{a = 1,\cdots,F} \;  ,  \; \mathcal{F}^{*}=\left(S^{i}\right)_{i = 1,\cdots,E} \; , \; \mathcal{P}^{*}=\left(P^{\alpha}\right)_{\alpha = 1,\cdots,V} \, ,
\eea
in such a way that:
\bea
\label{interdual}
P_{\mu} \odot x^{\nu} = \delta_{\mu}^{\nu} \; \; , \; \; S_{a} \odot e^{b} = \delta_{a}^{b} \; \; , \; \; e_{i} \odot S^{j} = \delta_{i}^{j} \; \; , \; \; x_{\alpha} \odot P^{\beta} = \delta_{\alpha}^{\beta} \, ,
\eea
with $\odot$ denoting the intersection number in $M$. The decompositions $\mathcal{C}$ and $\mathcal{C}^{*}$ are naturally endowed with the structure of abelian graded groups.

Let us list some important construction and properties that these dual decompositions yield.

{\it (1) Boundary operator.} We provide $\mathcal{C}$ and $\mathcal{C}^{*}$ with boundary operators $\oppartial$ and $\oppartial^\ast$ such that:
\bea
\label{boundary}
\left\{ \begin{aligned}
& \oppartial P_{\mu}  = \oppartial_{\mu}^{b} S_{b} \\
& \oppartial S_{a}  = \oppartial_{a}^{j} e_{j} \\
& \oppartial e_{i}  = \oppartial_{i}^{\beta} x^{\beta} \\
& \oppartial x_{\alpha}  = 0 \\
\end{aligned} \right.
\; \; \; \; \mbox{ and } \; \; \; \;
\left\{ \begin{aligned}
& \oppartial^\ast P^{\alpha} = {\oppartial^\ast}^{\alpha}_{j} S^{j} \\
& \oppartial^\ast S^{i} = {\oppartial^\ast}^{i}_{b} e^{b} \\
& \oppartial^\ast e^{a} = {\oppartial^\ast}^{a}_{\nu} x^{\nu} \\
& \oppartial^\ast x^{\mu} = 0 \\
\end{aligned}  \right. \, ,
\eea
with:
\bea
\label{null}
\oppartial \circ \oppartial = 0 = \oppartial^\ast \circ \oppartial^\ast \, ,
\eea
all matrix elements of $\oppartial$ and $\oppartial^\ast$ being integers. By introducing the matrix notation:
\bea
\label{Matrixoppartial}
\begin{gathered}
  \oppartial  = \left( {\begin{array}{*{20}{c}}
  0&\vline & 0&\vline & 0&\vline & 0 \\
 \hline
  {\left( {{\oppartial _{\left( 3 \right)}}{{^a}_\mu }} \right)_{a,\mu}}&\vline & 0&\vline & 0&\vline & 0 \\
\hline
  0&\vline & {\left( {{\oppartial _{\left( 2 \right)}}{{^i}_a}} \right)_{i,a}}&\vline & 0&\vline & 0 \\
\hline
  0&\vline & 0&\vline & {\left( {{\oppartial _{\left( 1 \right)}}{{^\alpha }_i}} \right)_{\alpha,i}}&\vline & 0
\end{array}} \right) \hfill \\
\end{gathered}
\eea
we have:
\bea
\label{dualpartial}
{\left( {{\oppartial^\ast_{\left( 3 \right)}} = \oppartial _{\left( 1 \right)}^\dag } \right)_{E \times V}} \; , \; {\left( {{\oppartial^\ast_{\left( 2 \right)}} = \oppartial _{\left( 2 \right)}^\dag } \right)_{F \times E}} \; ,\; {\left( {{\oppartial^\ast_{\left( 1 \right)}} = \oppartial _{\left( 3 \right)}^\dag } \right)_{P \times F}} \, ,
\eea
The boundary operators $\oppartial$ and $\oppartial^\ast$ turn $\mathcal{C}$ and $\mathcal{C}^{*}$ into differential groups \cite{McL} thus yielding homology groups $H_\bullet(\mathcal{C})$ and $H_\bullet(\mathcal{C}^\ast)$. We will always assume that the decomposition $\mathcal{C}$ is \textbf{good}, which means that:
\bea
\label{gooddecompo}
H_\bullet(\mathcal{C}) \simeq H_\bullet(M) \, .
\eea
By construction the dual decomposition $\mathcal{C}^\ast$ is good too.
\vspace{2mm}

{\it (2) Cochains and differentials.} Relations \eqref{interdual} lead to the following correspondences:
\bea
\label{assocochain1}
\left\{ \begin{aligned}
& P^{\alpha} \rightarrow \hat{P}^{\alpha} \in \Hom{\mathcal{V}}{\Z} \equiv C^0_{\mathcal{C}} \; \; / \; \; \hat{P}^{\alpha}(x_\beta) = \delta^\alpha_\beta \\
& S^{i} \rightarrow \hat{S}^{i} \in \Hom{\mathcal{E}}{\Z} \equiv C^1_{\mathcal{C}} \; \; / \; \; \hat{S}^{i}(e_j) = \delta^i_j \\
& e^{a} \rightarrow \hat{e}^{a} \in \Hom{\mathcal{F}}{\Z} \equiv C^2_{\mathcal{C}} \; \; / \; \; \hat{e}^{a}(S_b) = \delta^a_b \\
& x^{\mu} \rightarrow \hat{x}^{\mu} \in \Hom{\mathcal{P}}{\Z} \equiv C^3_{\mathcal{C}} \; \; / \; \; \hat{x}^{\mu}(P_\nu) = \delta^\mu_\nu \\
\end{aligned}  \right. \, ,
\eea
and once $\left(\mathcal{C}^{*}\right)^{*}$ has been canonically identified with $\mathcal{C}$ the following additional correspondences can be done:
\bea
\label{assocochain2}
\left\{ \begin{aligned}
& P_{\mu} \rightarrow \hat{P}_{\mu} \in \Hom{\mathcal{V}^{*}}{\Z} \equiv C^0_{\mathcal{C}^\ast} \; \; / \; \; \hat{P}_{\mu}(x^{\nu}) = \delta_{\mu}^{\nu} \\
& S_{a} \rightarrow \hat{S}_{a} \in \Hom{\mathcal{E}^{*}}{\Z} \equiv C^1_{\mathcal{C}^\ast} \; \; / \; \; \hat{S}_{a} (e^{b}) = \delta_{a}^{b} \\
& e_{i} \rightarrow \hat{e}_{i} \in \Hom{\mathcal{F}^{*}}{\Z} \equiv C^2_{\mathcal{C}^\ast} \; \; / \; \; \hat{e}_{i} (S^{j}) = \delta_{i}^{j}\\
& x_{\alpha}\rightarrow \hat{x}_{\alpha} \in \Hom{\mathcal{P}^{*}}{\Z} \equiv C^3_{\mathcal{C}^\ast} \; \; / \; \; \hat{x}_{\alpha}(P^\beta) = \delta_\alpha^\beta \\
\end{aligned} \right. \, .
\eea
A cochain of $\mathcal{C}$ and $\mathcal{C}^{*}$ is then a linear combination of these fundamental cochains.

We write $C^\bullet_{\mathcal{C}}$ (resp. $C^\bullet_{\mathcal{C}^\ast}$) the graded group of cochains of $\mathcal{C}$ (resp. $\mathcal{C}^\ast$). We turn $C^\bullet_{\mathcal{C}}$ (resp. $C^\bullet_{\mathcal{C}^\ast}$) into a differential group by endowing it with the endomorphism $d : C^\bullet_{\mathcal{C}} \rightarrow C^\bullet_{\mathcal{C}}$ (resp. $d^\ast : C^\bullet_{\mathcal{C}^\ast} \rightarrow C^\bullet_{\mathcal{C}^\ast}$) defined by:
\bea
\label{diff}
\begin{gathered}
\forall \hat{u} \in C^\bullet_{\mathcal{C}}, \quad d \circ \hat{u} = \hat{u} \circ \oppartial \\
\left( \mbox{resp.} \; \; \; \; \forall \hat{v} \in C^\bullet_{\mathcal{C}^\ast}, \quad d^\ast \circ \hat{v} = \hat{v} \circ \oppartial^\ast \right)
\end{gathered} \, .
\eea
Since the decomposition $\mathcal{C}$ is good the cohomology groups of $\left(C^\bullet_{\mathcal{C}},d \right)$ and $\left(C^\bullet_{\mathcal{C}^\ast},d^\ast \right)$ coincide with the ones of $M$. With respect to expression \eqref{Matrixoppartial} we have the matrix relations:
\bea
\label{derivresppartial}
{\left( {{d_{\left( 0 \right)}} = \oppartial _{\left( 1 \right)}^\dag } \right)_{E \times V}} \; , \; {\left( {{d_{\left( 1 \right)}} = \oppartial _{\left( 2 \right)}^\dag } \right)_{F \times E}} \; ,\; {\left( {{d_{\left( 2 \right)}} = \oppartial _{\left( 3 \right)}^\dag } \right)_{P \times F}} \, ,
\eea
and:
\bea
\label{deriv*resppartial}
{\left( {d_{\left( 0 \right)}^ *  = d_{\left( 2 \right)}^\dag  = {\oppartial _{\left( 3 \right)}}} \right)_{F \times P}} \; ,\; {\left( {d_{\left( 1 \right)}^ *  = d_{\left( 1 \right)}^\dag  = {\oppartial _{\left( 2 \right)}}} \right)_{E \times F}} \; ,\; {\left( {d_{\left( 2 \right)}^ *  = d_{\left( 0 \right)}^\dag  = {\oppartial _{\left( 1 \right)}}} \right)_{V \times E}} \, .
\eea
\vspace{2mm}

{\it (3) Cap and cup.} The symmetric non-degenerate pairings defined by:
\bea
\label{defpairings}
\left\{ \begin{aligned}
& \left\langle\hat{P}_{\mu},\hat{x}^{\nu}\right\rangle \equiv \hat{P}_{\mu}\left(x^{\nu}\right) = P_{\mu}\odot x^{\nu} = \delta_{\mu}^{\nu} \\
& \left\langle\hat{S}_{a},\hat{e}^{b}\right\rangle \equiv \hat{S}_{a}\left(e^{b}\right) =  S_{a}\odot e^{b} = \delta_{a}^{b} \\
& \left\langle\hat{e}_{i},\hat{S}^{j}\right\rangle \equiv \hat{e}_{i}\left(S^{j}\right) = e_{i} \odot S^{j} = \delta_{i}^{j} \\
& \left\langle\hat{x}_{\alpha},\hat{P}^{\beta}\right\rangle \equiv  \hat{x}_{\alpha}\left(P^{\beta}\right) = x_{\alpha}\odot P^{\beta} = \delta_{\alpha}^{\beta} \\
\end{aligned} \right.\, ,
\eea
yield the following cap products:
\bea
\label{capprod}
\left\{ \begin{aligned}
& M \smallfrown \hat{P}_{\mu} = P_{\mu} \\
& M \smallfrown \hat{S}_{a} = S_{a} \\
& M \smallfrown \hat{e}_{i} = e_{i} \\
& M \smallfrown \hat{x}_{\alpha} = x_{\alpha} \\
\end{aligned} \right.
\; \; \mbox{ and } \; \;
\left\{ \begin{aligned}
& M \smallfrown \hat{P}^{\alpha} = P^\alpha \\
& M \smallfrown \hat{S}^{i} = S^{i} \\
& M \smallfrown \hat{e}^{a} = e^{a} \\
& M \smallfrown \hat{x}^{\mu} = x^{\mu} \\
\end{aligned} \right. \, .
\eea
These relations are nothing but \textbf{Poincaré duality}. For instance, for a $1$-chain $c = c^{i} e_{i}$ then its Poincaré dual is just $\hat{c} = c^{i} \hat{e}_{i} \in C^2_{\mathcal{C}^\ast}$. Note that we start with a chain in $\mathcal{C}$ and end with a cochain in $\mathcal{C}^\ast$.

The cup products associated to the previous cap products are:
\bea
\label{copairings}
\left\{ \begin{aligned}
& \left(\hat{P}_{\mu} \smallsmile \hat{x}^{\nu} \right) (M) \equiv \hat{P}_{\mu} (M \smallfrown \hat{x}^{\nu}) = \hat{x}^{\nu} (M \smallfrown \hat{P}_{\mu}) = \hat{P}_{\mu} (x^{\nu}) = \delta_{\mu}^{\nu} \\
& \left( \hat{S}_{a} \smallsmile \hat{e}^{b}\right) (M) \equiv \hat{S}_{a}(M \smallfrown \hat{e}^{b}) = \hat{e}^{b} (M \smallfrown \hat{S}_{a}) = \hat{S}_{a}(e^{b}) = \delta_{a}^{b} \\
& \left( \hat{e}_{i} \smallsmile \hat{S}^{j}\right) (M) \equiv \hat{e}_{i} (M \smallfrown \hat{S}^{j}) = \hat{S}^{j} (M \smallfrown \hat{e}_{i}) = \hat{e}_{i}(S^{j}) = \delta_{i}^{j} \\
& \left( \hat{x}_{\alpha} \smallsmile \hat{P}^{\beta}\right) (M) \equiv  \hat{x}_{\alpha} (M \smallfrown \hat{P}^{\beta}) = \hat{P}^{\beta} (M \smallfrown \hat{x}_{\alpha}) = \hat{x}_{\alpha} (P^{\beta}) = \delta_{\alpha}^{\beta} \\
\end{aligned} \right. \, .
\eea
\vspace{2mm}

{\it (4) Labelings and gaugings.} The previous construction extends to $\Z_N$-valued cochains of $\mathcal{C}$ and $\mathcal{C}^{*}$ the differential groups of which are denoted $C^{N,\bullet}_{\mathcal{C}}$ and $C^{N,\bullet}_{\mathcal{C}^\ast}$. In the context of Turaev-Viro theory \cite{TV,BK,MT} elements of $C^{N,1}_{\mathcal{C}}$ (resp. $C^{N,1}_{\mathcal{C}^\ast}$) are called $\Z_N$ \textbf{labelings} of $\mathcal{C}$ (resp. $\mathcal{C}^\ast$) whereas elements of $C^{N,0}_{\mathcal{C}}$ (resp. $C^{N,0}_{\mathcal{C}^\ast}$) are called $\Z_N$ \textbf{gaugings} of $\mathcal{C}$ (resp. $\mathcal{C}^\ast$).

By construction, the differential of a $\Z_N$ gauging is a $\Z_N$ labeling.

Let us consider $\hat{l} \in C^{N,1}_{\mathcal{C}}$ and $\hat{m} \in C^{N,1}_{\mathcal{C}^\ast}$ such that:
\bea
\label{decompolabel}
\hat{l} = l_{i} \hat{S}^{i} \; \;
 \; \; \mbox{and} \; \; \; \; \hat{m} = m^{a} \hat{S}_{a} \, ,
\eea
with $l_{i}, m^{a} \in \Z_N$. The $2$-cochains $d\hat{l} \in C^{N,2}_{\mathcal{C}}$ and $d^\ast \hat{m} \in C^{N,2}_{\mathcal{C}^\ast}$ are defined by equation \eqref{diff} which gives:
\bea
\label{diffl}
d \hat{l} = \left(l_{i}\oppartial^{i}_{a}\right) \hat{e}^{a} = (d \hat{l})_a \hat{e}^{a}  \; \;
 \; \; \mbox{and} \; \; \; \; d^\ast \hat{m} = \left({m^{a} \oppartial^\ast}^{a}_{i}\right) \hat{e}^{i} = \left(d^\ast \hat{m}\right)_i \hat{e}^{i} \, .
\eea
Note that $(d \hat{l})_a \in \Z_N$ since $\oppartial^{i}_{a} \in \Z$ and $\Z_N$ is a $\Z$-module. Thanks to the ring structure of $\Z_N$ we can extend the cup products \eqref{capprod} to $\Z_N$-valued cochains. In particular we have:
\bea
\label{cuplabels}
\left( \hat{m} \smallsmile d\hat{l} \right) (M) = m^{a} \, (d\hat{l})_b \left( \hat{S}_a \smallsmile \hat{e}^b \right) (M) =  m^{a} (dl)_{a} \, .
\eea

As $C^{N,1}_{\mathcal{C}} = \Hom{\mathcal{E}}{\Z_{N}} \simeq \Z_{N}^E$ and $C^{N,1}_{\mathcal{C}^\ast} = \Hom{\mathcal{E}^{*}}{\Z_{N}} \simeq \Z_{N}^F$ we introduce the canonical bijections:
\bea
\begin{aligned}
\label{vectorcorresp}
\hat{l} = l_{i} \hat{S}^{i} \in C^{N,1}_{\mathcal{C}} & \longrightarrow \textbf{l} = (l_{i})_{i=1,\cdots,E} \in \Z^{E}_{N} \\
\hat{m} = m_{i} \hat{S}^{i} \in C^{N,1}_{\mathcal{C}^\ast} & \longrightarrow \textbf{m} = (m^{a})_{a=1,\cdots,F} \in \Z^{E}_{N}
\end{aligned} \, ,
\eea
so that we have:
\bea
\begin{aligned}
\label{dvectorcorresp}
d \hat{l} \in C^{N,2}_{\mathcal{C}} & \longrightarrow d \textbf{l}  = \left((dl)_{a}\right)_{a=1,\cdots,F} \in \Z^{F}_{N} \\
d^\ast \hat{m} \in C^{N,2}_{\mathcal{C}^\ast} & \longrightarrow d^\ast \textbf{m} = \left((d^\ast m)^{i}\right)_{i=1,\cdots,F} \in \Z^{E}_{N}
\end{aligned} \, .
\eea
Poincaré duality implies that a chain has the same components than its Poincaré dual regardless of the fact that these components are taken in $\Z$ or $\Z_N$. For instance the Poincaré dual of $c = c^i e_i$ is the $2$-cochain $\hat{c} = c^i \hat{e}_i$ since $M \smallfrown (c^i \hat{e}_i) = c^i (M \smallfrown \hat{e}_i) = c^i e_i$.

Using correspondences \eqref{vectorcorresp} and \eqref{dvectorcorresp}, we can rewrite equation \eqref{cuplabels} as:
\bea
\left( \hat{m} \smallsmile d\hat{l} \right) (M) = \textbf{m} \cdot d \textbf{l} = \left( d^* \hat{m} \smallsmile \hat{l} \right) (M)  \, ,
\eea
where the $\cdot$ denotes the euclidian scalar product.
\vspace{2mm}

{\it (5) Holonomy.} If $z = z^{i} e_{i}$ is a $1$-cycle of $\mathcal{C}$ then for any $\hat{l} = l_{i} \hat{S}^{i} \in C^{N,1}_{\mathcal{C}}$ we have:
\bea
\label{eq3}
\hat{l}(z) = \left( l_{i}\hat{S}^{i}\right)(z^{j} e_{j}) = l_{i} z^{i} = \textbf{l} \cdot \textbf{z} = (\hat{l} \smallsmile \hat{z})(M) \, ,
\eea
with $\textbf{z} = (z^i)_{i=1,\cdots,E} \in \Z^{E}$ and $\hat{z} = z^{i} \hat{e}_{i}$ the Poincaré dual of $z$. In the same way if $z^{*} = z^\ast_a e^a$ is a cycle of $\mathcal{C}^{*}$ then for any cochain $\hat{m} = m^{a} \hat{S}_{a}  \in C^{N,1}_{\mathcal{C}^\ast}$ we have:
\bea
\hat{m} (z^\ast) = \left( m^{a} \hat{S}_{a}\right) (z^\ast_{b} e^{b}) = m^{a} z^\ast_{a}  = \textbf{m} \cdot \textbf{z}^\ast  = (\hat{m} \smallsmile \hat{z}^\ast)(M) \, ,
\eea
with $\textbf{z}^\ast = (z^\ast_a)_{a=1,\cdots,F} \in \Z^{F}$ and $\hat{z} = z^\ast_{a} \hat{e}^{a}$ the Poincaré dual of $z^\ast$.

For any $\hat{l} \in C^{N,1}_{\mathcal{C}}$ and $\hat{m} \in C^{N,1}_{\mathcal{C}^\ast}$ the cochains $\hat{l}/N$ and $\hat{m}/N$ are $\mathbb{R}/\mathbb{Z}$-valued. Their holonomies are:
\bea
e^{2 i \pi \frac{\hat{l}}{N}(z)} = e^{ \frac{2 i \pi}{N} \hat{l}(z)} = e^{ \frac{2 i \pi}{N} \textbf{l} \cdot \textbf{z}}  \; \; \; \; \mbox{and} \; \; \; \; e^{2 i \pi \frac{\hat{m}}{N}(z^*)} = e^{ \frac{2 i \pi}{N} \hat{m}(z^*)} = e^{ \frac{2 i \pi}{N} \textbf{m} \cdot \textbf{z}^\ast} \, ,
\eea
where $z$ is a cycle of $\mathcal{C}^\ast$ and $z^*$ a cycle of $\mathcal{C}^\ast$. In particular this means that $\hat{l}/N$ and $\hat{m}/N$ are $\Z_N$-connections on $\mathcal{C}^{*}$ and $\mathcal{C}^{*}$ respectively.

\subsection{Abelian TV partition function and observables}

Let us assume that $M$ is provided with a cellular decompostion $\mathcal{C}$ as described in (3.1). In \cite{MT} we presented an abelian version of the TV invariant whose expression in $\mathcal{C}$ is:
\bea
\label{upsilondef}
\Upsilon_{N} = \frac{1}{N^{V-1}} \sum\limits_{\hat{l} \in C^{N,1}_{\mathcal{C}}} \left(\prod\limits_{S \in \mathcal{F}} \delta^{\left[N\right]}_{\Sigma^{\hat{l}}_{S}} \right)
\eea
where $\Sigma^{\hat{l}}_{S} = d \hat{l} (S)$. Using correspondences \eqref{vectorcorresp} and \eqref{dvectorcorresp} we can write $\Upsilon_{N}$ as:
\bea
\label{uspilonalt}
\Upsilon_{N} = \frac{1}{N^{V-1}} \sum\limits_{\textbf{l} \in \Z_{N}^E} \delta^{\left[N\right]}_{d\textbf{l}} \, ,
\eea
and by transforming Kronecker symbols into complex exponentials we obtain:
\bea
\label{TVpartition}
\Upsilon_{N} = \frac{1}{N^{F+V-1}}
\sum\limits_{{\textbf{m}}\in\Z^{F}_{N}}
\sum\limits_{{\textbf{l}}\in\Z^{E}_{N}}
e^{\frac{2i\pi}{N}{\textbf{m}}\cdot d{\textbf{l}}} = \frac{1}{N^{F+V-1}}
\sum\limits_{{\hat{m}} \in C^{N,1}_{\mathcal{C}^\ast}}
\sum\limits_{{\hat{l}} \in C^{N,1}_{\mathcal{C}}}
e^{\frac{2i\pi}{N} \left( {\hat{m}} \smallsmile d{\hat{l}} \right) (M)} \, .
\eea
We can also write $\Upsilon_{N}$ in terms of the $\Z_N$ connections of $\mathcal{C}$ and $\mathcal{C}^\ast$ as:
\bea
\label{TVpartitconnec}
\Upsilon_{N} = \frac{1}{N^{F+V-1}}
\sum\limits_{{\hat{m}} \in C^{N,1}_{\mathcal{C}^\ast}}
\sum\limits_{{\hat{l}} \in C^{N,1}_{\mathcal{C}}}
e^{2 i \pi N  \left( \left(\frac{\hat{m}}{N}\right) \smallsmile d \left(\frac{\hat{l}}{N}\right) \right) (M)} \, ,
\eea
Remembering that the cup product of cochains is the equivalent of the wedge product of forms and that locally $A \star B = A \wedge dB$, we can notice the similarity of expression \eqref{TVpartitconnec} with the BF partition function \eqref{BFparitiondef}.

Under the form \eqref{TVpartitconnec} the invariant $\Upsilon_{N}$ appears like a discretisation of the abelian BF partition function for the action $\left(\frac{\hat{m}}{N}\right) \smallsmile d \left(\frac{\hat{l}}{N}\right)$. The fields appearing in this action are $\Z_N$ connections of $\mathcal{C}$ and $\mathcal{C}^\ast$ and the coupling constant is $N$ like in BF. We hence refer to \eqref{TVpartitconnec} as the $\Z_N$ TV theory.

After these remarks it seems natural to consider as relevant observables of the $\Z_N$ TV theory the $\Z_{N}$-holonomies:
\bea
e^{\frac{2i\pi}{N}{ {\textbf{l}} \cdot \textbf{z}}_{1} } \; \; \; \; \mbox{ and } \; \; \; \; e^{\frac{2i\pi}{N}{\textbf{m}}\cdot{\textbf{z}}_{2}} \, ,
\eea
with ${\textbf{z}}_{1} \in \Z^E$ representing a cycle of $\mathcal{C}$ and ${\textbf{z}}_{2}  \in \Z^F$ a cycle of $\mathcal{C}^{*}$. The expectation values of these observables with respect to $\Upsilon_{N}$ are obviously defined as:
\bea
\label{vevz1z2}
\left\langle\left\langle z_{1}, z_{2}\right\rangle\right\rangle_{TV_{N}} = \frac{1}{N^{F+V-1}}
\sum\limits_{{\textbf{m}}\in\Z^{F}_{N}}
\sum\limits_{{\textbf{l}}\in\Z^{E}_{N}}
e^{\frac{2i\pi}{N}{\textbf{m}} \cdot d{\textbf{l}}}
e^{\frac{2i\pi}{N}{{\textbf{l}}\cdot \textbf{z}}_{1}}
e^{\frac{2i\pi}{N}{\textbf{m}} \cdot {\textbf{z}}_{2}} \, ,
\eea
or in terms of the $\Z_N$ connections of $\mathcal{C}$ and $\mathcal{C}^\ast$ as:
\bea
\label{vevz1z2}
\left\langle\left\langle z_{1}, z_{2}\right\rangle\right\rangle_{TV_{N}} = \frac{1}{N^{F+V-1}}
\sum\limits_{{\hat{m}} \in C^{N,1}_{\mathcal{C}^\ast}}
\sum\limits_{{\hat{l}} \in C^{N,1}_{\mathcal{C}}}
e^{2 i \pi \left\{ N \left(\left(\frac{\hat{m}}{N}\right) \smallsmile d \left(\frac{\hat{l}}{N}\right)\right)(M)
+ \left(\frac{\hat{l}}{N}\right) (z_{1})
+ \left(\frac{\hat{m}}{N}\right) (z_{2}) \right\} } \, .
\eea
This last expression has to be compared with the expectation value of holonomies in the usual $U(1)$ BF theory but also with expression \eqref{vevBF}. We can introduce the Poincaré duals $\hat{z}_1$ and $\hat{z}_2$ of ${z}_1$ and ${z}_2$ thus getting:
\bea
\label{vevz1z3}
\left\langle\left\langle z_{1}, z_{2}\right\rangle\right\rangle_{TV_{N}} = \frac{1}{N^{F+V-1}}
\sum\limits_{{\hat{m}} \in C^{N,1}_{\mathcal{C}^\ast}}
\sum\limits_{{\hat{l}} \in C^{N,1}_{\mathcal{C}}}
e^{2 i \pi \left\{ \left( N \left(\frac{\hat{m}}{N}\right) \smallsmile d \left(\frac{\hat{l}}{N}\right) + \left(\frac{\hat{l}}{N}\right) \smallsmile \hat{z}_{1} + \left(\frac{\hat{m}}{N}\right) \smallsmile \hat{z}_{2} \right) (M) \right\} } \, .
\eea
In the $U(1)$ BF theory this corresponds to write holonomies with the use of the de Rham currents of ${z}_1$ and ${z}_2$, and in the DB framework of section 2 to relation \eqref{holonomy}.

\subsection{Gauge fixing procedures}

Once the TV invariant \eqref{upsilondef} as been written under the form of the partition function \eqref{TVpartitconnec} we can wonder whether a gauge fixing procedure could be used instead of the normalisation factor $1/N^{V-1}$. Before discussing this let us make a remark concerning the expression of the TV partition function. By construction \cite{TV} it depends on $M$ and not on the chosen cellular decomposition of $M$. Hence instead of using the cellular decomposition $\mathcal{C}$ we can use the dual one $\mathcal{C}^\ast$. This means that we have:
\bea
\label{comparedTV}
\frac{1}{N^{V-1}} \sum\limits_{\hat{l} \in \C^{N,1}_{\mathcal{C}}} \delta^{\left[N\right]}_{d\hat{l}} = \frac{1}{N^{V^\ast-1}} \sum\limits_{\hat{m} \in \C^{N,1}_{\mathcal{C}^\ast}} \delta^{\left[N\right]}_{d^\ast \hat{m}} \, .
\eea
Even by noticing that $V^\ast = P$ this equality does not seem trivial. However if we use the exponential form of the Kronecker symbol $\delta^{[N]}$ to rewrite this relation we obtain:
\bea
\label{uspilonalt}
\frac{1}{N^{F+V-1}}
\sum\limits_{{\hat{m}} \in C^{N,1}_{\mathcal{C}^\ast}}
\sum\limits_{{\hat{l}} \in C^{N,1}_{\mathcal{C}}}
e^{\frac{2i\pi}{N} \left( {\hat{m}} \smallsmile d{\hat{l}} \right) (M)}
= \frac{1}{N^{F^\ast +V^\ast -1}} \sum\limits_{{\hat{m}} \in C^{N,1}_{\mathcal{C}^\ast}}
\sum\limits_{{\hat{l}} \in C^{N,1}_{\mathcal{C}}}
e^{\frac{2i\pi}{N} \left( {\hat{l}} \smallsmile d^\ast {\hat{m}} \right) (M)}  \, ,
\eea
and since $V^\ast = P$, $F^\ast = E$  and $\left( {\hat{l}} \smallsmile d^\ast {\hat{m}} \right) (M) = \left( {\hat{m}} \smallsmile d{\hat{l}} \right) (M)$ we have just to compare $1/N^{F+V-1}$ with $1/N^{E+P-1}$. It turns out that the Euler characteristic of $M$ is zero which implies that $V-E+F-P = 0$ and hence that $E+P-1 = F+V-1$ thus showing that \eqref{uspilonalt} and hence \eqref{comparedTV} hold true.

In \cite{FL} a geometrical gauge fixing procedure is proposed in the non-abelian context. To apply this procedure to our abelian case we consider an oriented spanning tree $T$ in $\mathcal{C}$ rooted at a vertex $x_0$ of $\mathcal{C}$. Such a graph always exists thanks to $M$ connectedness, reaches any vertex of $\mathcal{C}$ and does not contain any cycle. The orientation of $T$ is defined by going from the root $x_0$, which is the only vertex with no incoming edge, to any vertex of $\mathcal{C}$. This orientation induces a canonical orientation of the edges of $T$ so that for any $e \in T$ we write $\partial e = t(e) - s(e)$, where $t(e)$ (resp. $s(e)$) denotes the target (resp. the source) of $e$ with respect to its canonical orientation.

The gauge fixing procedure is then to restrict the sum over labelings which defines the TV invariant \eqref{upsilondef} of $M$ to the labelings $\hat{l} \in C^{N,1}_{\mathcal{C}}$ which satisfy:
\bea
\label{treegaugefix}
\forall e \in T, \, \, \hat{l}(e) = 0  \, .
\eea
For such a gauge fixed labeling $\hat{l}$ the gauge transformed labeling $\hat{l} + d \hat{\mu}$, with $\hat{\mu} \in C^{N,0}_{\mathcal{C}}$ satisfies:
\bea
\label{gaugetranstree}
\forall e \in T, \, \, \left( \hat{l} + d \hat{\mu} \right) (e) = \hat{\mu}(\partial e) = \hat{\mu}(t(e) - s(e)) = \hat{\mu}(t(e)) - \hat{\mu}(s(e)) \, .
\eea
On the one hand every vertex of $\mathcal{C}$ belongs to $T$ and on the other hand $x_0$ is the root of $T$ hence we have:
\bea
\left( \hat{l} + d \hat{\mu} \right) (e) = 0 \, \Leftrightarrow \, \hat{\mu}(x) = \hat{\mu}(x_0) \, ,
\eea
which means that $\hat{\mu}$ is a constant gauging. Hence the geometrical gauge fixing selects one and only one representative in each cohomology class of $\mathbb{Z}_N$ cocycles of $C^{N,1}_{\mathcal{C}}$. This coincides with the result of \cite{MT} where the partition function was normalised by the quotient of the set of gaugings by the set of constant gaugings thus yielding the normalisation factor $1/N^{V-1}$. The construction is independent of the root $x_0$.

The ``spanning tree" gauge fixing we just described can be seen as a homotopic gauge in the following sense: consider a neighbourhood of the tree $T$. This defines a contractible open set of $M$ with origin $x_0$, the contraction being done along the edges of $T$ until we reach $x_0$. The first and second homology and cohomology groups of this open set are trivial. This means that the restriction of a closed labeling of $\mathcal{C}$ is necessarily trivial, that is to say a gauging, hence gauge condition \eqref{treegaugefix}. What is remarkable is that the gauge fixing constraint applied in this open set is enough to gauge fix all the closed labelings on $\mathcal{C}$. Actually we can shrink the original decomposition along $T$ thus getting a new cellular decomposition with only one vertex, $x_0$, and which provides the same TV invariant as the original decomposition. This reduced decomposition of $M$ has only cyclic edges based at $x_0$. If we denote by $\mathcal{C}_T$ the reduced decomposition of $\mathcal{C}$ with respect to $T$ we have:
\bea
\label{reducedTV}
\frac{1}{N^{V-1}} \sum\limits_{\hat{l} \in \C^{N,1}_{\mathcal{C}}} \delta^{\left[N\right]}_{d\hat{l}} = \sum\limits_{\hat{l} \in \C^{N,1}_{\mathcal{C}_T^\ast}} \delta^{\left[N\right]}_{d\hat{l}} \, .
\eea
Let us note that this gauge fixing procedure is quite unusual since it is not a constraint on labelings -- i.e. fields -- of $\mathcal{C}$ but rather a change of cellular decomposition for $M$. This is why it was referred to as a geometrical gauge fixing.
\vspace{2mm}

With the previous geometrical gauge fixing we do not really need expression \eqref{TVpartition} of the TV partition function. By considering the TV action $\left( {\hat{l}} \smallsmile d^\ast {\hat{m}} \right) (M)$ appearing in \eqref{TVpartition} we can think about some other gauge fixing procedures inspired by what is usually done in Gauge Field Theory. The first example that comes to mind is that of the covariant (or Lorentz) gauge which in our discrete context takes the form:
\bea
\label{gaugecov}
d^\ast {}^h \hat{l} = 0 \quad [N] \, ,
\eea
where ${}^h \hat{l}$ is the Hodge dual of $\hat{l} = l_i \hat{S}^i$, Hodge duality being defined in $C^{N,1}_{\mathcal{C}}$ by:
\bea
\label{Hodgeduality}
{}^h\!\hat{S}^i = \delta^{ij} \hat{e}_j \in C^{N,2}_{\mathcal{C}^\ast} \, .
\eea
We want to compare expression \eqref{uspilonalt} with the supposedly gauge fixed one:
\bea
\label{covfixedpart}
\sum\limits_{\hat{l} \in C^{N,1}_{\mathcal{C}}} \delta^{\left[N\right]}_{d\hat{l}} \delta^{\left[N\right]}_{d^\ast {}^h \hat{l}} \, ,
\eea
or rather expression \eqref{TVpartition} with:
\bea
\frac{1}{N^{F+V-1}}
\sum\limits_{{\hat{m}} \in C^{N,1}_{\mathcal{C}^\ast}}
\sum\limits_{{\hat{l}} \in C^{N,1}_{\mathcal{C}}}
\sum_{\hat{\lambda} \in C^{N,3}_{\mathcal{C}^\ast}}
e^{\frac{2i\pi}{N} \left\{ \left( {\hat{m}} \smallsmile d{\hat{l}} \right) (M) + \left( \hat{\lambda} \smallsmile d^h \hat{l} \right)(M) \right\}} \, ,
\eea
where $d^h = {}^h d^\ast {}^h$. Unfortunately we are faced with several difficulties.
First of all the covariant gauge fixing procedure is usually done on differential forms which are by definition real valued. In other words from Hodge decomposition theorem we know that in the cohomology class of a \textbf{real} $1$-cocycle $r$ there is a unique co-closed representative, that is to say a real $1$-cocycle $\hat{l}$ such that $d^\ast {}^h \hat{l} = 0$. However real cohomology forgets about torsion hence it is hopeless to try to impose \eqref{gaugecov} on torsion cocycles. Even if $M$ has no torsion a $\Z$-valued cohomology class does not necessarily have a $\Z$-valued co-closed representative, and dealing with $\Z_N$-valued cocycles does not improve the situation.

Let us assume for a moment that $M$ is torsionless and such that each $\Z_N$-valued cohomology class have a representative which fulfills \eqref{gaugecov}. The cellular decomposition introduces possible degeneracies in this gauge fixing procedure. Indeed if a closed labeling $\hat{l} \in C^{N,1}_{\mathcal{C}}$ fulfills \eqref{gaugecov} then the gauge transformed closed labeling $\hat{l} + d \hat{\mu}$ fulfills it too if and only if:
\bea
\label{Diophantgauge}
d \circ d^h \hat{\mu} = \Delta \hat{\mu} = N \hat{\rho} \, ,
\eea
for some $\rho \in C^{N,0}_{\mathcal{C}}$. However gaugings $\hat{\mu}$ such that
\bea
\label{dipahconst}
d \hat{\mu} = N \hat{\omega} \, ,
\eea
has to be excluded since they do not change $\mathbb{Z}_N$ labelings. Solving the diophantine system \eqref{Diophantgauge} while excluding solutions of \eqref{dipahconst} can be a tedious task. Fortunately there is a loophole if a Heegaard splitting $H \cup_\varphi H$ of $M$ is used. A cellular decomposition $\mathcal{C}$ of the Riemann surface $\Sigma = \partial H$ compatible with the diffeomorphism $\varphi$ -- which means that $\varphi(\mathcal{C})$ is also a decomposition of $\Sigma$ -- canonically induces a cellular decomposition $\mathcal{C}_\varphi$ of $M$. Remarkably a dual decomposition $\mathcal{C}^\ast_\varphi$ of $\mathcal{C}_\varphi$ contains only two vertices. Taking into account the remarks made at the beginning of this subsection we find that:
\bea
\label{uspilonaltHeeg}
\Upsilon_{N} = \frac{1}{N^{E+1}} \sum\limits_{{\hat{m}} \in C^{N,1}_{\mathcal{C}^\ast_\varphi}}
\sum\limits_{{\hat{l}} \in C^{N,1}_{\mathcal{C}_\varphi}}
e^{\frac{2i\pi}{N} \left( {\hat{l}} \smallsmile d^\ast {\hat{m}} \right) (M)}  \, .
\eea
Trying to replace the normalisation factor $1/N$ in \eqref{uspilonaltHeeg} by using the covariant gauge fixing \eqref{gaugecov} leads us to consider:
\bea
\label{uspilonaltgfixed}
\frac{1}{N^{2+E}} \sum\limits_{{\hat{m}} \in C^{N,1}_{\mathcal{C}^\ast_\varphi}}
\sum\limits_{{\hat{l}} \in C^{N,1}_{\mathcal{C}_\varphi}} \sum_{\hat{\lambda} \in C^{N,3}_{\mathcal{C}_\varphi}}
e^{\frac{2i\pi}{N} \left\{ \left( {\hat{l}} \smallsmile d^\ast {\hat{m}} \right)(M) + \left(\hat{\lambda} \smallsmile d^\ast_h \hat{m}\right)(M) \right\} }  \, ,
\eea
with $d^\ast_h = {}^h d {}^h$ and $P = 2$. The degeneracy of the gauge constraint is now much easier to study as $\oppartial_{(3)} (P_1 +P_2) = \oppartial_{(3)} M = 0$ and hence $\oppartial_{(3)} P_2 = - \oppartial_{(3)} P_1$. Thus the matrix of $\oppartial_{(3)}$ has the simple form:
\bea
\label{oppart3Heeg}
\left( {\begin{array}{*{20}{c}}
  {{\varepsilon _1}}&{ - {\varepsilon _1}} \\
  {{\varepsilon _2}}&{ - {\varepsilon _2}} \\
   \vdots & \vdots  \\
  {{\varepsilon _F}}&{ - {\varepsilon _F}}
\end{array}} \right) `, ,
\eea
with $\epsilon_i = 0, \pm 1$ for $i =1, \cdots , F$, and the matrix representing ${\Delta^\ast} = {}^h d_{(2)} {}^h d_{(0)}^\ast$ is:
\bea
\label{deltadual}
{\Delta^\ast} = \oppartial_{(3)}^\dag \oppartial_{(3)} =  \left( {\begin{array}{*{20}{c}}
  n&{ - n} \\
  { - n}&n
\end{array}} \right) = n\left( {\begin{array}{*{20}{c}}
  1&{ - 1} \\
  { - 1}&1
\end{array}} \right)
\eea
where $n = \sum_i \epsilon^2_i$ is the number of common faces of $P_1$ and $P_2$, or equivalently the number of edges joining the two points of the dual cellular decomposition ${\mathcal{C}^\ast_\varphi}$. Equation \eqref{Diophantgauge} now reads:
\bea
\label{dualDiophantgauge}
\Delta^\ast \hat{\mu} = N \hat{\rho} = \left( {\begin{array}{*{20}{c}}
  n&{ - n} \\
  { - n}&n
\end{array}} \right)\left( {\begin{array}{*{20}{c}}
  {{\mu _1}} \\
  {{\mu _2}}
\end{array}} \right) = N\left( {\begin{array}{*{20}{c}}
  {{\rho _1}} \\
  {{\rho _2}}
\end{array}} \right) \, .
\eea
There are $\pgcd{N}{n} = k$ non-trivial -- i.e. such that $d_{(0)}^\ast \hat{\mu} = \oppartial_{(3)} \hat{\mu} \neq N \hat{\omega}$ -- solutions of this system. These are the degeneracies of the covariant gauge fixing and hence we have:
\bea
\label{uspilonalttruegfixed}
\Upsilon_{N} = \frac{1}{k N^{2+E}} \sum\limits_{{\hat{m}} \in C^{N,1}_{\mathcal{C}^\ast_\varphi}}
\sum\limits_{{\hat{l}} \in C^{N,1}_{\mathcal{C}_\varphi}} \sum_{\hat{\lambda} \in C^{N,3}_{\mathcal{C}_\varphi}}
e^{\frac{2i\pi}{N} \left\{ \left( {\hat{l}} \smallsmile d^\ast {\hat{m}} \right)(M) + \left(\hat{\lambda} \smallsmile d^\ast_h \hat{m}\right)(M) \right\} }  \, ,
\eea
that is to say:
\bea
\label{uspilonalttruegfixedend}
\Upsilon_{N} = \frac{1}{k} \sum\limits_{\hat{m} \in C^{N,1}_{\mathcal{C}^\ast_\varphi}} \delta^{\left[N\right]}_{d^\ast\hat{m}} \, \delta^{\left[N\right]}_{d^\ast_h \hat{m}}  \, .
\eea
An example where the covariant gauge fixing procedure just described can be applied is provided by Heegaard splittings $H \cup_\varphi H$ with $\varphi = Id_{\Sigma}$. Such a splitting defines a manifold $M$ such that $H_1(M) = \Z^g$, with $g$ the genus of $\Sigma$. The case of $S^1 \times S^2$ presented in subection 5.2 is of this kind. In any event, as natural as it seems to be the covariant gauge fixing procedure turns out to be much less effective than the geometrical one in the context of $U(1)$ TV theory.

As $\Z_N$ holonomies are gauge invariant, once the partition function has been properly gauge fixed, expectation values of these holonomies can be computed in the chosen gauge.

\section{Reciprocity formula}

We now show the main result of this paper:
\bea
\left\langle\left\langle z_{1}, z_{2}\right\rangle\right\rangle_{TV_{N}}
= \frac{N^{b_{1}}}{p_{1} \cdots p_{n}} \left\langle\left\langle z_{1} , z_{2}\right\rangle\right\rangle_{BF_{N}}
\eea

Since any cellular cycle is a cycle in $M$ whereas the converse is not true it seems natural to start from the TV theory. So let $z_{1} = z_{1}^{0}+z_{1}^{f}+z_{1}^{\tau} \in \mathcal{C}$ and $z_{2} = z_{2}^{0}+z_{2}^{f}+z_{2}^{\tau} \in \mathcal{C}^\ast$ be two cellular cycles. They yield the following expectation value:
\bea
\label{eq4}
\begin{aligned}
\left\langle\left\langle z_{1}, z_{2}\right\rangle\right\rangle_{TV_{N}}
& = \frac{1}{N^{F+V-1}}
\displaystyle\sum\limits_{{\textbf{m}}\in\Z^{F}_{N}}
\displaystyle\sum\limits_{{\textbf{l}}\in\Z^{E}_{N}}
e^{\frac{2i\pi}{N}{\textbf{m}} \cdot d{\textbf{l}}}
e^{\frac{2i\pi}{N}{{\textbf{l}} \cdot \textbf{z}}_{1}}
e^{\frac{2i\pi}{N}{\textbf{m}}\cdot{\textbf{z}}_{2}}\\
& = \frac{1}{N^{V-1}}
\displaystyle\sum\limits_{{\textbf{l}} \in \Z^{E}_{N}}
e^{\frac{2i\pi}{N}{{\textbf{l}} \cdot \textbf{z}}_{1}}
\delta^{\left[N\right]}_{d{\textbf{l}}+{\textbf{z}}_{2}}
\end{aligned} \, .
\eea
The sum over ${\textbf{m}}$ yields the constraint:
\bea
\label{constraint1}
d{\textbf{l}}+{\textbf{z}}_{2} = - N {\textbf{u}} \, ,
\eea
for some ${\textbf{u}} \in \Z^F$. The minus sign appearing in the right hand side is only here for later convenience. Constraint \eqref{constraint1} implies that the cycle $z_{2}$ of $\mathcal{C}^\ast$ can be seen -- through Poincaré duality -- as a $\Z_{N}$-coboundary. Moreover since $z_{2}$ is a cycle this same constraint also implies that:
\bea
d{\textbf{u}}=0 \, ,
\eea
which states that ${\textbf{u}}$ represents a $2$-cocycle. Hence we deduce that equation \eqref{constraint1} does not admit any solution if $z^{f}_{2}$ is not $0$ modulo $N$. The same reasoning applies to $z_1$ when factorizing out $\textbf{l}$ instead of $\textbf{m}$. Therefore we have
\bea
\left\langle\left\langle z_{1}, z_{2}\right\rangle\right\rangle_{TV_{N}}
= \delta^{\left[N\right]}_{\textbf{z}^{f}_{1}}\delta^{\left[N\right]}_{\textbf{z}^{f}_{2}} \left\langle\left\langle z^{0}_{1} + z^{\tau}_{1} , z^{0}_{2} + z^{\tau}_{2} \right\rangle\right\rangle_{TV_{N}}
\, ,
\eea
as with BF.

Consider two cycles $z'_{1}=z_{1}^{0}+z_{1}^{\tau} \in \mathcal{C}$ and $z'_{2}=z_{2}^{0}+z_{2}^{\tau} \in \mathcal{C}^\ast$ of order $p_{1}$ and $p_{2}$ respectively, and hence without any free part. Then there exists a $2$-chain $\Sigma_{1} \in \mathcal{C}$ such that:
\bea
p_{1}z'_{1}=\partial\Sigma_{1} \, ,
\eea
or equivalently for the vector ${\boldsymbol{\sigma}}_{1} \in \Z^F$ representing the Poincaré dual of $\Sigma_{1}$ such that:
\bea
p_{1}\textbf{z}'_{1}=d{\boldsymbol{\sigma}}_{1} \, .
\eea
The quantity:
\bea
\boldsymbol{\sigma}_{1}\cdot d{\textbf{l}} \, ,
\eea
represents the intersection number of $\Sigma_{1}$ with the boundary whose Poincaré dual is represented by $d{\textbf{l}}$. Constraint \eqref{constraint1} then yields:
\bea
{\boldsymbol{\sigma}}_{1}\cdot d{\textbf{l}}
= - {\boldsymbol{\sigma}}_{1}\cdot\left(N{\textbf{u}} + \textbf{z}'_{2}\right)
= - N {\boldsymbol{\sigma}}_{1}\cdot{\textbf{u}}-{\boldsymbol{\sigma}}_{1}\cdot\textbf{z}'_{2} \, .
\eea
Due to the symmetry property of the intersection we have:
\bea
{\boldsymbol{\sigma}}_{1}\cdot d{\textbf{l}}
= d{\boldsymbol{\sigma}}_{1}\cdot{\textbf{l}}
= p_{1} \; \textbf{z}'_{1} \cdot {\textbf{l}} \, ,
\eea
which gives:
\bea
\label{firststep}
\textbf{z}'_{1}\cdot{\textbf{l}}
= - \frac{N{\boldsymbol{\sigma}}_{1} \cdot {\textbf{u}}}{p_{1}}
- \frac{{\boldsymbol{\sigma}}_{1}\cdot\textbf{z}'_{2}}{p_{1}} \, .
\eea
The construction above allows to associate an element ${\textbf{u}} \in \Z^F$ to an element ${\textbf{l}} \in \Z^E$. We need to determine the degeneracy of this pairing when trying to use it in order to relate the sum over $\textbf{m}$ and $\textbf{l}$ in \eqref{eq4} with the sum over  $\textbf{u}$ and $\textbf{v}$ in \eqref{finaldiscBF}. For this purpose we set:
\bea
S = \left\lbrace \textbf{l} \in \Z^{E} \; | \; \exists \textbf{u} \in \Z^{F} \; | \; d \textbf{l} + \textbf{z}'_{2} = - N \textbf{u}\right\rbrace
\eea
and
\bea
S' = \left\lbrace \textbf{u} \in \Z^{F} \; | \; \exists \textbf{l} \in \Z^{E} \; | \; d \textbf{l} + \textbf{z}_{2} = - N \textbf{u} \right\rbrace
\eea
Thus, the set of summation in our computation are $\slfrac{S}{N\Z}$ and $\slfrac{S'}{\image d}$.

We now use the following lemma that is proven in appendix:
\begin{lem}
\bea
\slfrac{S'}{\image d} \simeq {{\left(\slfrac{S}{N\Z}\right)} \over {\left(\slfrac{\noyau d}{N\Z}\right)}} \, .
\eea
\end{lem}

Thus our degeneracy factor is $|Ker d/N\Z|$. Since $H^1(M) \simeq H_2(M) $ is free, we have $b_{1}$ independent directions in this set ($b_{1}$ being the first Betti number) to which we can add an amiguity $d\chi = \chi_{0}-\chi$, with $\chi_{0}$ corresponding to an arbitrary vertex. Hence there remains $V-1$ possibilities for $\chi$ in order to get a non-zero $d\chi$. All the elements of $|Ker d/N\Z|$ having coefficients in $\Z_{N}$ we get:
\bea
\abs{\slfrac{\noyau d}{N\Z}} = N^{b_{1}+V-1} \,  ,
\eea
and therefore:
\bea
\label{demoTVBF}
\begin{aligned}
\left\langle\left\langle z'_{1}, z'_{2}\right\rangle\right\rangle_{TV_{N}}
& = \frac{1}{N^{V-1}}
\sum\limits_{{\textbf{l}}\in\Z^{E}_{N}}
e^{\frac{2i\pi}{N} {\textbf{l}} \cdot \textbf{z}'_{1}}
\delta^{\left[N\right]}_{d{\textbf{l}} + \textbf{z}'_{2}} \\
& = \frac{1}{N^{V-1}}
\sum\limits_{{\textbf{l}}\in\slfrac{S}{N\Z}}
e^{\frac{2i\pi}{N}{\textbf{l}} \cdot \textbf{z}'_{1}} \\
& = N^{b_{1}}
\sum\limits_{{\textbf{u}}\in\slfrac{S'}{\image d}}
e^{\frac{- 2i\pi}{N}\left(\frac{N{\boldsymbol{\sigma}}_{1}\cdot{\textbf{u}}}{p_{1}}
+\frac{{\boldsymbol{\sigma}}_{1}\cdot\textbf{z}'_{2}}{p_{1}}\right)} \\
& = N^{b_{1}} e^{-\frac{2i\pi}{N}\frac{{\boldsymbol{\sigma}}_{1}\cdot\textbf{z}'_{2}}{p_{1}}}
\sum\limits_{{\textbf{u}}\in\slfrac{S'}{\image d}}
e^{-2i\pi\frac{{\boldsymbol{\sigma}}_{1}\cdot{\textbf{u}}}{p_{1}}} \\
& = N^{b_{1}} e^{-\frac{2i\pi}{N}\link{z'_{1}}{z'_{2}}}
\sum\limits_{{\textbf{u}}\in\slfrac{S'}{\image d}}
e^{-2i\pi\link{z'_{1}}{u}} \, ,
\end{aligned}
\eea
where $z'_{1}$, $z'_{2}$ and $u$ are the cycles associated respectively to $\textbf{z}'_{1}$, $\textbf{z}'_{2}$ and ${\textbf{u}}$. Since $\link{z'_{1}}{u}\egzz Q\left({\textbf{n}}_{1},{\textbf{u}}\right)$, where ${\textbf{n}}_{1}$ is the cohomology class of $z'_{1}$, we can write:
\bea
\label{demoTVBF2}
\begin{aligned}
\left\langle\left\langle z'_{1}, z'_{2}\right\rangle\right\rangle_{TV_{N}}
& = N^{b_{1}} e^{-\frac{2i\pi}{N}\link{z'_{1}}{z'_{2}}}
\sum\limits_{{\textbf{u}}\in\slfrac{S'}{\image d}}
e^{-2i\pi Q\left({\textbf{n}}_{1},{\textbf{u}}\right)}\\
& = N^{b_{1}} e^{-\frac{2i\pi}{N}\link{z'_{1}}{z'_{2}}}
\sum\limits_{{\textbf{u}}\in T_{1}}
e^{-2i\pi Q\left({\textbf{n}}_{1},{\textbf{u}}\right)}
\delta_{- N {\textbf{u}} - {\textbf{n}}_{2},0} \, .
\end{aligned}
\eea
Since $Q$ is a non-degenerate quadratic form on $T_1(M)$ we can use it to dualise the Kronecker symbol, thus getting:
\bea
\label{demoTVBF3}
\begin{aligned}
\left\langle\left\langle z'_{1}, z'_{2}\right\rangle\right\rangle_{TV_{N}}
& = \frac{N^{b_{1}}}{p_{1}\cdots p_{n}} e^{-\frac{2i\pi}{N}\link{z'_{1}}{z'_{2}}}
\sum\limits_{{\textbf{u}} , {\textbf{v}} \in T_{1}}
e^{-2i\pi Q\left({\textbf{n}}_{1},{\textbf{u}}\right)}
e^{-2i\pi Q\left(N{\textbf{u}}+{\textbf{n}}_{2},{\textbf{v}}\right)} \\
& = \frac{N^{b_{1}}}{p_{1}\cdots p_{n}} e^{-\frac{2i\pi}{N}\link{z'_{1}}{z'_{2}}}
\sum\limits_{{\textbf{u}} , {\textbf{v}} \in T_{1}}
e^{-2i\pi \left\{ N Q\left({\textbf{u}},{\textbf{v}}\right)
+ Q\left({\textbf{n}}_{1},{\textbf{u}}\right)
+ Q\left({\textbf{n}}_{2},{\textbf{v}}\right)\right\}} \, .
\end{aligned}
\eea
Hence we have shown that:
\bea
\label{finalresult}
\left\langle\left\langle z_{1} , z_{2} \right\rangle\right\rangle_{TV_{N}}
= \frac{N^{b_{1}}}{p_{1} \cdots p_{n}} \left\langle\left\langle z_{1} , z_{2} \right\rangle\right\rangle_{BF_{N}} \, ,
\eea
which is the reciprocity formula we were looking for. If $\exists i \in \intervalleentier{1}{n} \; | \; \pgcd{N}{p_{i}}\nmid {\textbf{n}}_{1}^{i} \mbox{ or } {\textbf{n}}_{2}^{i}$ then both sides of the equality vanish. Indeed, in this case, equation \eqref{constraint1} has no solution. The proportionality factor appearing in \eqref{finalresult} is closely related to the one appearing in the Deloup-Turaev reciprocity formula \cite{DT}, which in turn emerges as a Reshetekhin-Turaev surgery formula in the context of the $U(1)$ Chern-Simons theory \cite{GT3}.

Strictly speaking formula \eqref{finalresult} has a sense of reading -- from the left to the right -- since as already noticed not every cycle in $M$ is a cycle of the cellular decomposition, whereas the converse is always true. Let us note that in equations \eqref{demoTVBF3} it is the linking form $Q$ that was used to exponentiate $\delta_{- N {\textbf{u}} - {\textbf{n}}_{2},0}$ not just because it is non degenerate but also because it goes to homology classes and is defined by the linking number which itself appears in equation \eqref{firststep}.

\section{Examples}

For the following examples, we exploit the fact that a Heegaard splitting can lead to a good cellular decomposition. However this is far than being the only possibility. Our examples are lens spaces and thus admit a genus 1 decomposition that we consider here. To reconstruct the manifold with the diagrams given below, we first identify the left and right edge of each rectangle, to generate two solid cylinders, whose opposite faces are bounded by the upper and lower edge. Those faces are then identified for each solid cylinder, giving two solid tori. Finally, we identify the boundary of these two solid tori via the gluing rule $h$. The orientations can be complicated when considering a cycle passing through the (common) boundary of the solid cylinders. The doted lines appearing in the drawings below are not considered as elements of the cellular decomposition but are drawn only for the convenience of the representation. By convention $z_1$ denotes a $1$-cycle of $\mathcal{C}$ and $z_2$ a $1$-cycle of $\mathcal{C}^\ast$.

\subsection{$S^{3}$}

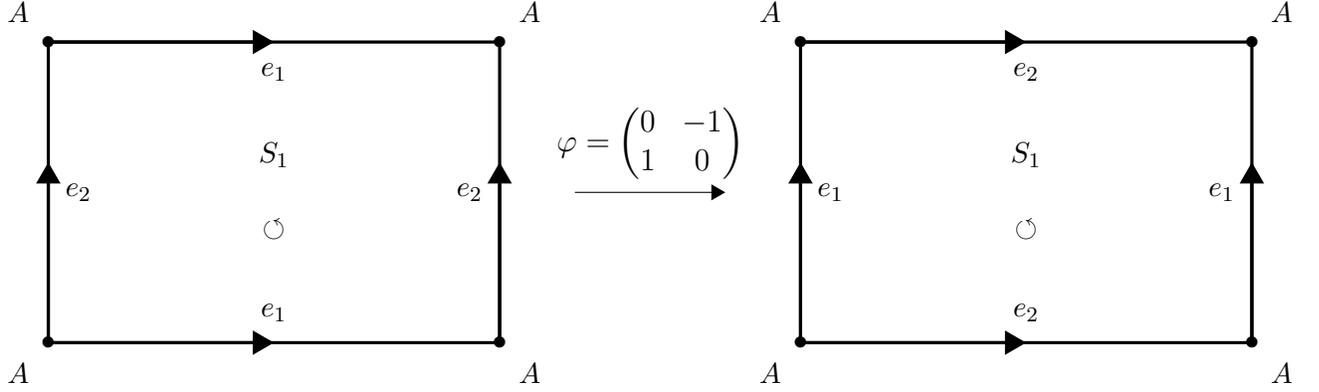
\begin{center}
\begin{tikzpicture}[scale=2.,>=triangle 60]
\draw[very thick] (0,0)-- (3,0);
\draw[very thick] (0,0)-- (0,2);
\draw[very thick] (0,2)-- (3,2);
\draw[very thick] (3,0)-- (3,2);

\draw (1.5,1.25) node {\small $S_{1}$};
\draw (1.5,0.75) node {\small $\circlearrowleft$};

\draw[very thick,->] (0,0)-- (1.5,0);
\draw[very thick,->] (0,2)-- (1.5,2);
\draw (1.5,1.8) node {\small $e_{1}$};
\draw (1.5,0.2) node {\small $e_{1}$};

\draw[very thick,->] (0,0)-- (0,1.2);
\draw[very thick,->] (3,0)-- (3,1.2);
\draw (0.2,1) node {\small $e_{2}$};
\draw (2.8,1) node {\small $e_{2}$};

\draw (0,0) node {\small $\bullet$};
\draw (-0.2,-0.2) node {\small $A$};
\draw (0,2) node {\small $\bullet$};
\draw (-0.2,2.2) node {\small $A$};
\draw (3,0) node {\small $\bullet$};
\draw (3.2,-0.2) node {\small $A$};
\draw (3,2) node {\small $\bullet$};
\draw (3.2,2.2) node {\small $A$};

\draw[->] (3.5,1) -- (4.5,1);
\draw (4,1) node[above]{$\varphi = \begin{pmatrix} 0 & -1 \\ 1 & 0 \end{pmatrix}$};

\draw[very thick] (5,0)-- (8,0);
\draw[very thick] (5,0)-- (5,2);
\draw[very thick] (5,2)-- (8,2);
\draw[very thick] (8,0)-- (8,2);

\draw (6.5,1.25) node {\small $S_{1}$};
\draw (6.5,0.75) node {\small $\circlearrowleft$};

\draw[very thick,->] (5,0)-- (5,1.2);
\draw[very thick,->] (8,0)-- (8,1.2);
\draw (5.2,1) node {\small $e_{1}$};
\draw (7.8,1) node {\small $e_{1}$};

\draw[very thick,->] (5,0)-- (6.5,0);
\draw[very thick,->] (5,2)-- (6.5,2);
\draw (6.5,1.8) node {\small $e_{2}$};
\draw (6.5,0.2) node {\small $e_{2}$};

\draw (5,0) node {\small $\bullet$};
\draw (4.8,-0.2) node {\small $A$};
\draw (5,2) node {\small $\bullet$};
\draw (4.8,2.2) node {\small $A$};
\draw (8,0) node {\small $\bullet$};
\draw (8.2,-0.2) node {\small $A$};
\draw (8,2) node {\small $\bullet$};
\draw (8.2,2.2) node {\small $A$};
\end{tikzpicture}
\end{center}

Here, $F=3$ (the faces $S_{2}$ and $S_{3}$ which do not appear on the diagram are the sections of the left and right solid cylinders whose boundaries are $e_{1}$ and $e_{2}$), $E=2$ and $V=1$. The operator $d_{(1)}$ being the transpose of the matrix giving the components of $\partial S_{i}$ in the basis $e_{j}$, we get:
\bea
d_{(1)} =
\begin{pmatrix}
0 & 0 \\ 1 & 0 \\ 0 & 1
\end{pmatrix}
\eea

With $z_{1} = e_{1} = \oppartial S_2$ and $z_{2} = e^{2}$ such that $\link{z_{1}}{z_{2}} = S_2 \odot e^{2} = 1$, we obtain:
\bea
\left\langle\left\langle z_{1}, z_{2}\right\rangle\right\rangle_{TV_{N}}
= e^{\frac{2i\pi}{N}}
= e^{\frac{2i\pi}{N}\link{z_{1}}{z_{2}}}
= \frac{N^{0}}{1}\left\langle\left\langle \gamma_{1}, \gamma_{2}\right\rangle\right\rangle_{BF_{N}} \, .
\eea
The general case arises by taking $z_{1} = n_1 e_{1}$ and $z_{2} = n_2 e^{2}$ so that $\link{z_{1}}{z_{2}} = n_1 n_2$. Note that no gauge fixing is required in this example.

\subsection{$S^{1} \times S^{2}$}

\begin{center}
\begin{tikzpicture}[scale=2.,>=triangle 60]
\draw[very thick] (0,0)-- (3,0);
\draw[dashed] (0,0)-- (0,2);
\draw[very thick] (0,2)-- (3,2);
\draw[dashed] (3,0)-- (3,2);
\draw[very thick] (0.75,0)-- (0.75,2);
\draw[very thick] (2.25,0)-- (2.25,2);

\draw (0.4,1.25) node {\small $S_{1}$};
\draw (0.4,0.75) node {\small $\circlearrowleft$};
\draw (1.5,1.25) node {\small $S_{2}$};
\draw (1.5,0.75) node {\small $\circlearrowleft$};
\draw (2.6,1.25) node {\small $S_{1}$};
\draw (2.6,0.75) node {\small $\circlearrowleft$};

\draw[very thick,->] (0,0)-- (0.4,0);
\draw[very thick,->] (0,2)-- (0.4,2);
\draw (0.4,1.8) node {\small $e_{1}$};
\draw (0.4,0.2) node {\small $e_{1}$};

\draw[very thick,->] (0,0)-- (1.6,0);
\draw[very thick,->] (0,2)-- (1.6,2);
\draw (1.6,1.8) node {\small $e_{2}$};
\draw (1.6,0.2) node {\small $e_{2}$};

\draw[very thick,->] (0,0)-- (2.7,0);
\draw[very thick,->] (0,2)-- (2.7,2);
\draw (2.7,1.8) node {\small $e_{3}$};
\draw (2.7,0.2) node {\small $e_{3}$};

\draw[very thick,->] (0.75,0)-- (0.75,1.1);
\draw (0.55,1) node {\small $e_{4}$};
\draw (0.95,1) node {\small $e_{4}$};

\draw[very thick,->] (2.25,0)-- (2.25,1.1);
\draw (2.05,1) node {\small $e_{5}$};
\draw (2.45,1) node {\small $e_{5}$};

\draw (0,0) node {\small $\bullet$};
\draw (-0.2,-0.2) node {\small $A$};
\draw (0,2) node {\small $\bullet$};
\draw (-0.2,2.2) node {\small $A$};
\draw (3,0) node {\small $\bullet$};
\draw (3.2,-0.2) node {\small $A$};
\draw (3,2) node {\small $\bullet$};
\draw (3.2,2.2) node {\small $A$};
\draw (0.75,0) node {\small $\bullet$};
\draw (0.75,-0.2) node {\small $B$};
\draw (0.75,2) node {\small $\bullet$};
\draw (0.75,2.2) node {\small $B$};
\draw (2.25,0) node {\small $\bullet$};
\draw (2.25,-0.2) node {\small $C$};
\draw (2.25,2) node {\small $\bullet$};
\draw (2.25,2.2) node {\small $C$};

\draw[->] (3.5,1) -- (4.5,1);
\draw (4,1) node[above]{$\varphi = \begin{pmatrix} 1 & 0 \\ 0 & 1 \end{pmatrix}$};

\draw[very thick] (5,0)-- (8,0);
\draw[dashed] (5,0)-- (5,2);
\draw[very thick] (5,2)-- (8,2);
\draw[dashed] (8,0)-- (8,2);
\draw[very thick] (5.75,0)-- (5.75,2);
\draw[very thick] (7.25,0)-- (7.25,2);

\draw (5.4,1.25) node {\small $S_{1}$};
\draw (5.4,0.75) node {\small $\circlearrowleft$};
\draw (6.5,1.25) node {\small $S_{2}$};
\draw (6.5,0.75) node {\small $\circlearrowleft$};
\draw (7.6,1.25) node {\small $S_{1}$};
\draw (7.6,0.75) node {\small $\circlearrowleft$};

\draw[very thick,->] (5,0)-- (5.4,0);
\draw[very thick,->] (5,2)-- (5.4,2);
\draw (5.4,1.8) node {\small $e_{1}$};
\draw (5.4,0.2) node {\small $e_{1}$};

\draw[very thick,->] (5,0)-- (6.6,0);
\draw[very thick,->] (5,2)-- (6.6,2);
\draw (6.6,1.8) node {\small $e_{2}$};
\draw (6.6,0.2) node {\small $e_{2}$};

\draw[very thick,->] (5,0)-- (7.7,0);
\draw[very thick,->] (5,2)-- (7.7,2);
\draw (7.7,1.8) node {\small $e_{3}$};
\draw (7.7,0.2) node {\small $e_{3}$};

\draw[very thick,->] (5.75,0)-- (5.75,1.1);
\draw (5.55,1) node {\small $e_{4}$};
\draw (5.95,1) node {\small $e_{4}$};

\draw[very thick,->] (7.25,0)-- (7.25,1.1);
\draw (7.05,1) node {\small $e_{5}$};
\draw (7.45,1) node {\small $e_{5}$};

\draw (5,0) node {\small $\bullet$};
\draw (4.8,-0.2) node {\small $A$};
\draw (5,2) node {\small $\bullet$};
\draw (4.8,2.2) node {\small $A$};
\draw (8,0) node {\small $\bullet$};
\draw (8.2,-0.2) node {\small $A$};
\draw (8,2) node {\small $\bullet$};
\draw (8.2,2.2) node {\small $A$};
\draw (5.75,0) node {\small $\bullet$};
\draw (5.75,-0.2) node {\small $B$};
\draw (5.75,2) node {\small $\bullet$};
\draw (5.75,2.2) node {\small $B$};
\draw (7.25,0) node {\small $\bullet$};
\draw (7.25,-0.2) node {\small $C$};
\draw (7.25,2) node {\small $\bullet$};
\draw (7.25,2.2) node {\small $C$};
\end{tikzpicture}
\end{center}

Here, $F=4$ (the faces $S_{3}$ and $S_{4}$ which do not appear on the diagram are the sections of the left and right solid cylinders whose boundaries are $e_{1}+e_{2}+e_{3}$), $E=5$ and $V=3$. The operator $d_{(1)}$ being the transpose of the matrix giving the components of $\partial S_{i}$ in the basis $e_{j}$, we get:
\bea
d_{(1)} =
\begin{pmatrix}
0 & 0 & 0 & 1 & -1 \\ 0 & 0 & 0 & -1 & 1 \\ 1 & 1 & 1 & 0 & 0 \\ 1 & 1 & 1 & 0 & 0
\end{pmatrix}
\eea

Computing the TV partition function, we obtain:
\bea
\Upsilon_{N} = \frac{1}{N^{F+V-1 = 6}}
\displaystyle\sum\limits_{{\textbf{l}}\in\Z_{N}^{E=5}}
\displaystyle\sum\limits_{{\textbf{m}}\in\Z_{N}^{F=4}}
e^{\frac{2i\pi}{N} {\textbf{m}}\cdot d{\textbf{l}}}
= N
= \frac{N^{1}}{1}Z_{BF_{N}}
\eea

Computing now the expectation value of $z_{1}=e_{1}+e_{2}+e_{3} = \oppartial S_3$ and $z_{2}=0$, we obtain:
\bea
\left\langle\left\langle z_{1}, z_{2}\right\rangle\right\rangle_{TV_{N}}
= \frac{1}{N^{6}}
\displaystyle\sum\limits_{{\textbf{l}}\in\Z_{N}^{5}}
\displaystyle\sum\limits_{{\textbf{m}}\in\Z_{N}^{4}}
e^{\frac{2i\pi}{N}\left({\textbf{m}}\cdot d{\textbf{l}}
+{\textbf{l}}\cdot{\textbf{z}_{1}}\right)}
= N
= \frac{N^{1}}{1}\left\langle\left\langle \gamma_{1}, \gamma_{2}\right\rangle\right\rangle_{BF_{N}}
\eea

With $z_{1}=e_{1}+e_{2}+e_{3} = \oppartial S_3$ and $z_{2}=e^{1}-e^{2}+e^{3}+e^{4}$ trivial such that $\link{z_{1}}{z_{2}} = S_3 \odot (e^{1}-e^{2}+e^{3}+e^{4}) = S_3 \odot e^{3} = 1$, we obtain:
\bea
\left\langle\left\langle z_{1}, z_{2}\right\rangle\right\rangle_{TV_{N}}
= N e^{\frac{2i\pi}{N}}
= N e^{\frac{2i\pi}{N}\link{z_{1}}{z_{2}}}
= \frac{N^{1}}{1}\left\langle\left\langle \gamma_{1}, \gamma_{2}\right\rangle\right\rangle_{BF_{N}}
\eea

With $z_{1}=e_{4}$ non-trivial and $z_{2}=0$, we obtain:
\bea
\left\langle\left\langle z_{1}, z_{2}\right\rangle\right\rangle_{TV_{N}}
=0
= \frac{N^{1}}{1}\left\langle\left\langle \gamma_{1}, \gamma_{2}\right\rangle\right\rangle_{BF_{N}}
\eea
as expected. It can  be checked that the covariant gauge fixing procedure can be applied in this example.

\subsection{${\R}P^{3}=L\left(2,1\right)$}

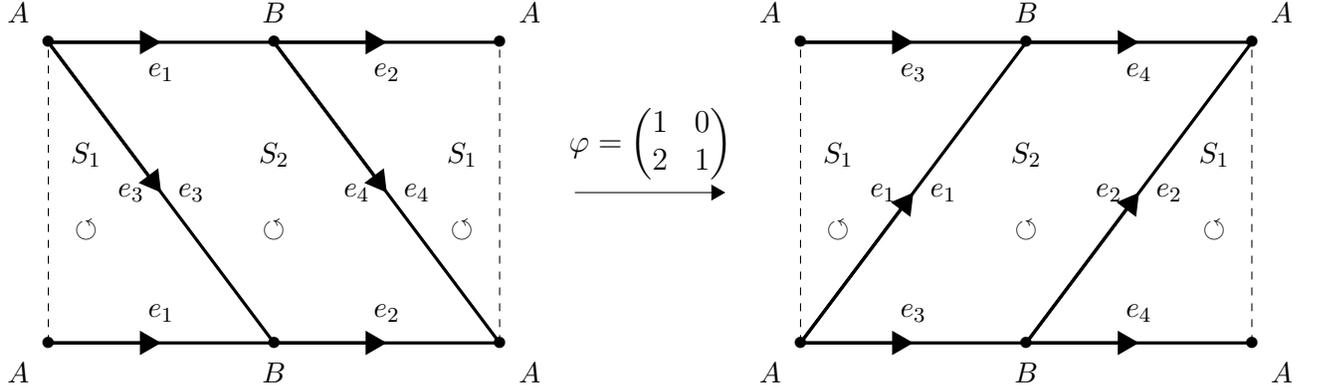
\begin{center}
\begin{tikzpicture}[scale=2.,>=triangle 60]
\draw[very thick] (0,0)-- (3,0);
\draw[dashed] (0,0)-- (0,2);
\draw[very thick] (0,2)-- (3,2);
\draw[dashed] (3,0)-- (3,2);
\draw[very thick] (0,2)-- (1.5,0);
\draw[very thick] (1.5,2)-- (3,0);

\draw (0.25,1.25) node {\small $S_{1}$};
\draw (0.25,0.75) node {\small $\circlearrowleft$};
\draw (1.5,1.25) node {\small $S_{2}$};
\draw (1.5,0.75) node {\small $\circlearrowleft$};
\draw (2.75,1.25) node {\small $S_{1}$};
\draw (2.75,0.75) node {\small $\circlearrowleft$};

\draw[very thick,->] (0,0)-- (0.75,0);
\draw[very thick,->] (0,2)-- (0.75,2);
\draw (0.75,1.8) node {\small $e_{1}$};
\draw (0.75,0.2) node {\small $e_{1}$};

\draw[very thick,->] (1.5,0)-- (2.25,0);
\draw[very thick,->] (1.5,2)-- (2.25,2);
\draw (2.25,1.8) node {\small $e_{2}$};
\draw (2.25,0.2) node {\small $e_{2}$};

\draw[very thick,->] (0,2)-- (0.75,1);
\draw (0.55,1) node {\small $e_{3}$};
\draw (0.95,1) node {\small $e_{3}$};

\draw[very thick,->] (1.5,2)-- (2.25,1);
\draw (2.05,1) node {\small $e_{4}$};
\draw (2.45,1) node {\small $e_{4}$};

\draw (0,0) node {\small $\bullet$};
\draw (-0.2,-0.2) node {\small $A$};
\draw (0,2) node {\small $\bullet$};
\draw (-0.2,2.2) node {\small $A$};
\draw (3,0) node {\small $\bullet$};
\draw (3.2,-0.2) node {\small $A$};
\draw (3,2) node {\small $\bullet$};
\draw (3.2,2.2) node {\small $A$};
\draw (1.5,0) node {\small $\bullet$};
\draw (1.5,-0.2) node {\small $B$};
\draw (1.5,2) node {\small $\bullet$};
\draw (1.5,2.2) node {\small $B$};

\draw[->] (3.5,1) -- (4.5,1);
\draw (4,1) node[above]{$\varphi = \begin{pmatrix} 1 & 0 \\ 2 & 1 \end{pmatrix}$};

\draw[very thick] (5,0)-- (8,0);
\draw[dashed] (5,0)-- (5,2);
\draw[very thick] (5,2)-- (8,2);
\draw[dashed] (8,0)-- (8,2);
\draw[very thick] (5,0)-- (6.5,2);
\draw[very thick] (6.5,0)-- (8,2);

\draw (5.25,1.25) node {\small $S_{1}$};
\draw (5.25,0.75) node {\small $\circlearrowleft$};
\draw (6.5,1.25) node {\small $S_{2}$};
\draw (6.5,0.75) node {\small $\circlearrowleft$};
\draw (7.75,1.25) node {\small $S_{1}$};
\draw (7.75,0.75) node {\small $\circlearrowleft$};

\draw[very thick,->] (5,0)-- (5.75,0);
\draw[very thick,->] (5,2)-- (5.75,2);
\draw (5.75,1.8) node {\small $e_{3}$};
\draw (5.75,0.2) node {\small $e_{3}$};

\draw[very thick,->] (6.5,0)-- (7.25,0);
\draw[very thick,->] (6.5,2)-- (7.25,2);
\draw (7.25,1.8) node {\small $e_{4}$};
\draw (7.25,0.2) node {\small $e_{4}$};

\draw[very thick,->] (5,0)-- (5.75,1);
\draw (5.55,1) node {\small $e_{1}$};
\draw (5.95,1) node {\small $e_{1}$};

\draw[very thick,->] (6.5,0)-- (7.25,1);
\draw (7.05,1) node {\small $e_{2}$};
\draw (7.45,1) node {\small $e_{2}$};

\draw (5,0) node {\small $\bullet$};
\draw (4.8,-0.2) node {\small $A$};
\draw (5,2) node {\small $\bullet$};
\draw (4.8,2.2) node {\small $A$};
\draw (8,0) node {\small $\bullet$};
\draw (8.2,-0.2) node {\small $A$};
\draw (8,2) node {\small $\bullet$};
\draw (8.2,2.2) node {\small $A$};
\draw (6.5,0) node {\small $\bullet$};
\draw (6.5,-0.2) node {\small $B$};
\draw (6.5,2) node {\small $\bullet$};
\draw (6.5,2.2) node {\small $B$};
\end{tikzpicture}
\end{center}

Here, $F=4$ (the faces $S_{3}$ and $S_{4}$ which do not appear on the diagram are the sections of the left and right solid cylinders whose boundaries are $e_{1}+e_{2}$ and $e_{3}+e_{4}$), $E=4$ and $V=2$. The operator $d_{(1)}$ being the transpose of the matrix giving the components of $\partial S_{i}$ in the basis $e_{j}$, we get:
\bea
d_{(1)} =
\begin{pmatrix}
1 & -1 & -1 & 1 \\ -1 & 1 & 1 & -1 \\ 1 & 1 & 0 & 0 \\ 0 & 0 & 1 & 1
\end{pmatrix}
\eea

For $z_{1}=e_{1}+e_{4}$ (with $2 z_1 = \oppartial (S_1 + S_3 + S_4)$) and $z_{2} = e^{3}+e^{4}$ trivial such that $\link{z_{1}}{z_{2}} = \frac{1}{2} \link{2z_{1}}{z_{2}} = \frac{1}{2} (S_1 + S_3 + S_4) \odot (e^{3}+e^{4}) = 1$, we obtain:
\bea
\left\langle\left\langle z_{1}, z_{2}\right\rangle\right\rangle_{TV_{N}}
= e^{-\frac{2i\pi}{N}} \left(1 - \delta^{[2]}_{N}\right)
= e^{-\frac{2i\pi}{N}\link{z_{1}}{z_{2}}} \left(1 - \delta^{[2]}_{N}\right)
= \frac{N^{0}}{2}\left\langle\left\langle \gamma_{1}, \gamma_{2}\right\rangle\right\rangle_{BF_{N}}
\eea
the cohomology class $\textbf{n}_{1}$ associated to $z_{1}$ being $1$ and $\textbf{n}_{2}$ associated to $z_{2}$ being $0$.

With $z_{1}=e_{1}+e_{4}$ and $z_{2} = e^{3}$ torsion such that $\frac{1}{2} \link{2z_{1}}{z_{2}} = \frac{1}{2} (S_1 + S_3 + S_4) \odot e^{3} = \frac{1}{2}$, we obtain:
\bea
\left\langle\left\langle z_{1}, z_{2}\right\rangle\right\rangle_{TV_{N}}
= -e^{-\frac{i\pi}{N}} \left(1 - \delta^{[2]}_{N}\right)
= -e^{-\frac{2i\pi}{N}\link{z_{1}}{z_{2}}} \left(1 - \delta^{[2]}_{N}\right)
= \frac{N^{0}}{2}\left\langle\left\langle \gamma_{1}, \gamma_{2}\right\rangle\right\rangle_{BF_{N}}
\eea
the cohomology class $\textbf{n}_{1}$ associated to $z_{1}$ being $1$ and $\textbf{n}_{2}$ associated to $z_{2}$ being $1$.

The presence of $\delta^{[2]}_{N}$ in the above expressions comes from the fact that for even $N$, $\pgcd{N}{p=2} = 2 \nmid \textbf{n}_{1}, \textbf{n}_{2}$ implying that \eqref{constraint1} has no solution. It can be checked at the level of the partition function that in this case the covariant gauge fixing does not apply properly since it produces a $\pgcd{N}{4}$ factor.

\section{Conclusion}
In this article we showed how the use of Deligne-Beilinson cohomology allows to prove that the $U(1)$ BF theory can be turned into a discrete $\mathbb{Z}_N$ BF theory without resorting to the usual guessworks of the non-abelian case. For instance all the sums occurring in the discrete theory are finite thanks to the emergence of $\mathbb{Z}_N$ as ``gauge" group whereas in the non-abelian case a Quantum Group is introduced as a way to regularise the infinite sums the non-abelian discrete BF theory yields. In addition it is only under this regularisation assumption that the non-abelian BF theory is related with a TV invariant whereas in the $U(1)$ this relation is proven too.
However it has to be stressed out that although the discrete BF action on $M$ is $\left( {\hat{l}} \smallsmile d^\ast {\hat{m}} \right) (M)$ the action of the corresponding $U(1)$ BF theory is NOT $\int_M B \wedge dA$ but $\int_M A \star B$. It's only on $S^3$ that $\int_M B \wedge dA$ becomes a possible expression for the $U(1)$ BF action since the set of $U(1)$-connections on $S^3$ can be identified with $\Omega^1(S^3)$ and the gauge group with $d \Omega^0(S^3)$ (see exact sequence \eqref{exactsequence1}).

Finally, all we have done in this article can be extended to connected, closed, smooth and oriented  manifold of dimension $m = 4l+3$ with configuration space of the BF theory being $H^{2l+1}_D(M) \times H^{2l+1}_D(M)$ instead of $H^1_D(M) \times H^1_D(M)$ and the one of TV being $C^{N,2l+1}_{\mathcal{C}} \times C^{N,2l+1}_{\mathcal{C}^\ast}$ instead of $C^{N,1}_{\mathcal{C}} \times C^{N,1}_{\mathcal{C}^\ast}$ for some cellular dual decompositions $\mathcal{C}$ and $\mathcal{C}^\ast$ of $M$.

\section{Appendix: Proof of the lemma}

We now want to prove the following:
\begin{lem}
\bea
\slfrac{S'}{\image d} \simeq {{\left(\slfrac{S}{N\Z}\right)} \over {\left(\slfrac{\noyau d}{N\Z}\right)}} \, .
\eea
\end{lem}
\vspace{2mm}

\begin{dem}
Let's consider:
\bea
\begin{array}{ccrcl}
\varphi &: & \slfrac{S'}{\image d} & \to & \slfrac{\left(\slfrac{S}{N\Z}\right)}{\left(\slfrac{\noyau d}{N\Z}\right)} \\
 & & \overline{\textbf{u}} & \mapsto & \overline{\overline{\textbf{l}}} \\
\end{array}
\eea
where we use bars to emphasise the fact that we work with classes in the appropriate quotient sets.

First we check that $\varphi$ is well-defined, that is to say
\bea
\overline{\textbf{u}}=\overline{\textbf{v}}\Rightarrow\varphi\left(\overline{\textbf{u}}\right)=\varphi\left(\overline{\textbf{v}}\right).
\eea
Indeed,
\bea
\overline{\textbf{u}} = \overline{\textbf{v}} \Rightarrow \overline{\textbf{u}} - \overline{\textbf{v}} = 0 \Rightarrow \overline{\textbf{u} - \textbf{v}} = 0 \, ,
\eea
which means by definition that:
\bea
\textbf{u} - \textbf{v} \in \image d \; \Rightarrow \; \exists \textbf{a} \in \Z^{E} \; | \; \textbf{u} - \textbf{v} = d \textbf{a} \, .
\eea
But
\bea
\textbf{u} \in S' \; \Rightarrow \;  \exists \textbf{l} \in \Z^{E} \; | \; d \textbf{l} + \textbf{z}'_{2} = - N \textbf{u} \, ,
\eea
and
\bea
\textbf{v} \in S' \; \Rightarrow \; \exists \textbf{m} \in \Z^{E} \; | \; d \textbf{m} + \textbf{z}'_{2} = - N \textbf{v} \, ,
\eea
(the minus sign in the right-hande side being purely conventional) thus
\bea
N \left(\textbf{u} - \textbf{v}\right) = N d \textbf{a} = - d \left(\textbf{l} - \textbf{m}\right)
\; \; \; \; \mbox{ or } \; \; \; \;
d \left(\textbf{l} - \textbf{m} + N \textbf{a}\right) = 0 \, ,
\eea
and so
\bea
\textbf{l} = \textbf{m} - N \textbf{a} + \boldsymbol{\xi} \, ,
\eea
with $\boldsymbol{\xi}\in\noyau d$, so
\bea
\overline{\textbf{l}}=\overline{\textbf{m}+\boldsymbol{\xi}} \, ,
\eea
and hence
\bea
\overline{\overline{\textbf{l}}} = \overline{\overline{\textbf{m}}} \; \; \Rightarrow \; \; \varphi\left(\overline{\textbf{u}}\right)=\varphi\left(\overline{\textbf{v}}\right).
\eea

Then, we note that, by construction, $\varphi$ is necessarily surjective. Thus we only need to prove that it is injective. For that we consider $\overline{\overline{\textbf{l}}}=\overline{\overline{\textbf{m}}}$, that is to say
\bea
\overline{\overline{\textbf{l}}} - \overline{\overline{\textbf{m}}} = \overline{\overline{\textbf{l} - \textbf{m}}} = 0 \; \; \Rightarrow \; \; \overline{\textbf{l }- \textbf{m}} = \overline{\boldsymbol{\xi}} \, ,
\eea
with $\overline{\boldsymbol{\xi}}\in\slfrac{\noyau d}{\Z_{N}}$ so
\bea
\exists \textbf{a} \in \Z^{E} \; | \; \textbf{l} - \textbf{m} = \boldsymbol{\xi} - N \textbf{a} \, .
\eea
But
\bea
\textbf{l} \in S \; \; \Rightarrow \; \; \exists \textbf{u} \in \Z^{F} \; | \; d \textbf{l} + \textbf{z}'_{2} = - N \textbf{u} \, ,
\eea
and
\bea
\textbf{m} \in S \; \; \Rightarrow \; \; \exists \textbf{v} \in \Z^{F} \; | \; d \textbf{m} + \textbf{z}'_{2} = - N \textbf{v} \, ,
\eea
thus
\bea
- N \textbf{u} = d \textbf{l} + \textbf{z}'_{2} = d \left(\textbf{m} - N \textbf{a} + \boldsymbol{\xi}\right) + \textbf{z}'_{2} = \left(d\textbf{m} + \textbf{z}'_{2}\right) - N d\textbf{a} = N \left(\textbf{v} - d \textbf{a}\right)
\eea
so
\bea
\textbf{u} = \textbf{v} - d\textbf{a}\Rightarrow\overline{\textbf{u}} = \overline{\textbf{v}} \, .
\eea

Hence, $\varphi$ is bijective.
\qed
\end{dem}

\vfill\eject

\end{document}